%% file: main.tex
\documentclass[10pt]{article}

\usepackage{fullpage}
\usepackage[utf8]{inputenc}
\usepackage[T1]{fontenc}
\usepackage{parskip}
\usepackage[numbers]{natbib}
\usepackage{hyperref}
\usepackage{url}
\usepackage{booktabs}
\usepackage{amsfonts}
\usepackage{nicefrac}
\usepackage{microtype} 
\usepackage{xcolor} 
\usepackage{amsmath}
\usepackage{amsthm}
\usepackage{bm}
\usepackage{cleveref}
\usepackage{tikz}
\usepackage{enumitem}
\usepackage[ruled]{algorithm2e}
\usepackage{setspace}
\usepackage{subcaption}
\usepackage{multirow}
\usepackage{autonum}
\usepackage[most]{tcolorbox}

\def\X{\mathcal{X}}
\def\Y{\mathcal{Y}}
\def\F{\mathcal{F}}
\def\B{F}
\def\A{\hat{\mu}}
\def\D{\operatorname{D}}
\def\G{\mathcal{G}}
\def\H{\mathcal{H}}
\def\Hess{\operatorname{Hess}}

\def\R{\mathbb{R}}
\def\P{\mathcal{P}}
\def\P2{\mathcal{P}_2}
\def\W2{\mathcal{W}_2}
\def\KL{\operatorname{KL}}
\def\E{\mathbb{E}}
\def\rv{\theta}

\theoremstyle{definition}
\newtheorem{example}{Example}
\newtheorem{remark}{Remark}
\SetKwInOut{Parameter}{Parameter}

\title{Wasserstein Gradient Boosting: A Framework for Distribution-Valued Supervised Learning}

\author{
	Takuo Matsubara\\
	The University of Edinburgh\\
	\texttt{takuo.matsubara@ed.ac.uk} \\
}

\date{}

\begin{document}
	
	\maketitle
	
	\begin{abstract}
		\noindent Gradient boosting is a sequential ensemble method that fits a new weaker learner to pseudo residuals at each iteration.
		We propose Wasserstein gradient boosting, a novel extension of gradient boosting that fits a new weak learner to alternative pseudo residuals that are Wasserstein gradients of loss functionals of probability distributions assigned at each input.
		It solves distribution-valued supervised learning, where the output values of the training dataset are probability distributions for each input.
		In classification and regression, a model typically returns, for each input, a point estimate of a parameter of a noise distribution specified for a response variable, such as the class probability parameter of a categorical distribution specified for a response label.
		A main application of Wasserstein gradient boosting in this paper is tree-based evidential learning, which returns a distributional estimate of the response parameter for each input.
		We empirically demonstrate the superior performance of the probabilistic prediction by Wasserstein gradient boosting in comparison with existing uncertainty quantification methods.
	\end{abstract}

	\section{Introduction} \label{sec:introduction}
	
	Gradient boosting is a celebrated machine learning algorithm that has achieved considerable success with tabular data \cite{Shwartz-Ziv2022}.
	Gradient boosting has been extensively used for point forecasts and probabilistic classification, yet a relatively small number of studies have been concerned with the predictive uncertainty of gradient boosting.
	Predictive uncertainty of machine learning models plays a growing role in today's real-world production systems \cite{Abdar2021}.
	It is vital for safety-critical systems, such as medical diagnoses \cite{Topol2019} and autonomous driving \cite{Sorin2020}, to assess the potential risk of their actions that partially or entirely rely on predictions from their models.
	Gradient boosting has already been applied in a diverse range of real-world applications, including click prediction \cite{Richardson2007}, ranking systems \cite{Burges2010}, scientific discovery \cite{Roe2005}, and data competition \cite{Bennett2007}.
	There is a pressing need for methodology to harness the power of gradient boosting to predictive uncertainty quantification.
	
	In classification and regression, we typically specify a noise distribution $p(y \mid \rv)$ of a response variable $y$ and use a model to return a point estimate $\rv(x)$ of the response parameter for each input $x$.
	In recent years, the importance of capturing uncertainty in the model output $\rv(x)$ has increasingly been emphasised \cite{Abdar2021}.
	A variety of approaches have been proposed to obtain a distributional estimate $p( \rv \mid x )$ of the response parameter for each input $x$ \cite[e.g.][]{Gal2016,Lakshminarayanan2017,Sensoy2018}.
	For example, Bayesian neural networks (BNNs) quantify uncertainty in network weights and propagate it to the space of network outputs.
	Marginalising the predictive distribution $p(y \mid \rv)$ over the distributional estimate $p( \rv \mid x )$ has been demonstrated to confer enhanced predictive accuracy and robustness against adversarial attacks \cite{Sensoy2018}.
	Furthermore, the dispersion of the distributional estimate has been used as a powerful indicator for out-of-distribution (OOD) detection \cite{Gawlikowski2023}.
	
	In this context, a line of research based on the concept of \emph{evidential learning} has recently gained significant attention \cite{Sensoy2018,Amini2020,Charpentier2020,Ulmer2023}.
	The idea can be interpreted as making use of the `individual-level' posterior $p(\rv \mid y_i )$ of the response parameter $\rv$ conditional on each individual datum $y_i$, which arises from the response-distribution likelihood $p(y_i \mid \rv )$ and a user-specified prior $p(\rv)$.
	If each individual-level posterior falls into a closed form characterised by some hyperparameter, neural networks can be trained by using the hyperparameter of the individual-level posterior as a target value to predict for each input.
	Outstanding performance and computational efficiency of the existing approaches have been delivered in a wide spectrum of engineering and medical applications \cite{Capellier2019,Hemmer2020,Soleimany2021,Gawlikowski2022}.
	However, the existing approaches are limited to neural networks and to the case where every individual-level posterior is in closed form so that the finite-dimensional hyperparameter can be predicted by proxy.
	In general, posterior distributions are known only up to their normalising constants and, therefore, require an approximation typically by particles \cite{Gelman2013}.

	Without closed-form expression, each individual-level posterior needs to be treated as an infinite-dimensional output for each input.
	This challenge poses the following fundamental question:
	
	\begin{quote}
	\begin{tcolorbox}[colback=red!2!white,colframe=red!75!black]
		Consider supervised learning whose outputs are probability distributions.
		Given a training set of input values and output distributions $\{ x_i, \mu_i \}_{i=1}^{D}$, can we build a model that receives an input $x$ and returns a \emph{nonparametric} prediction of the output distribution?
	\end{tcolorbox}
	\end{quote}
	
\begin{figure}[t]
		\centering
		\subcaptionbox{$0$ weak learner trained}{\includegraphics[width=0.315\textwidth]{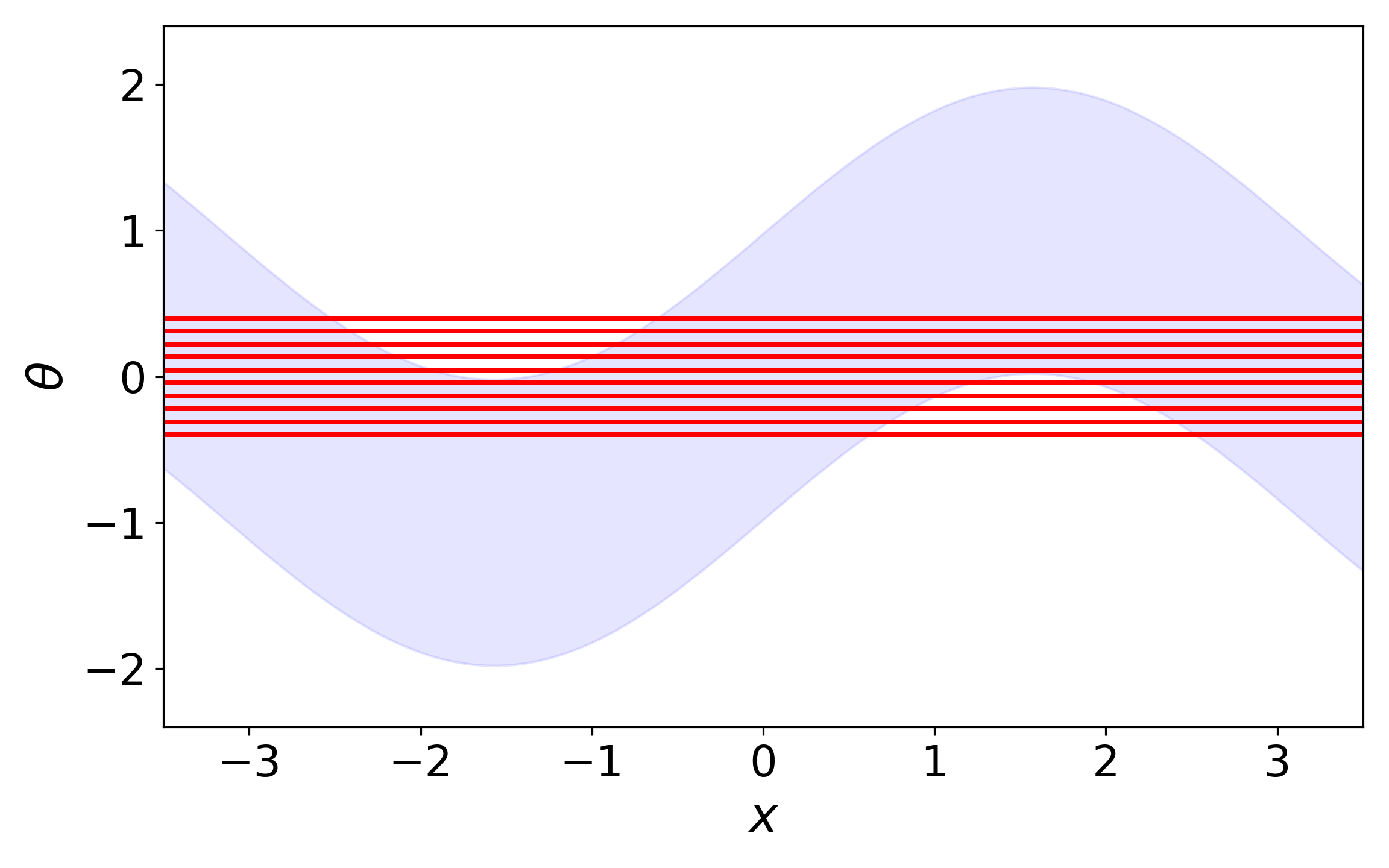}}
		\hfill
		\subcaptionbox{$15$ weak learners trained}{\includegraphics[width=0.315\textwidth]{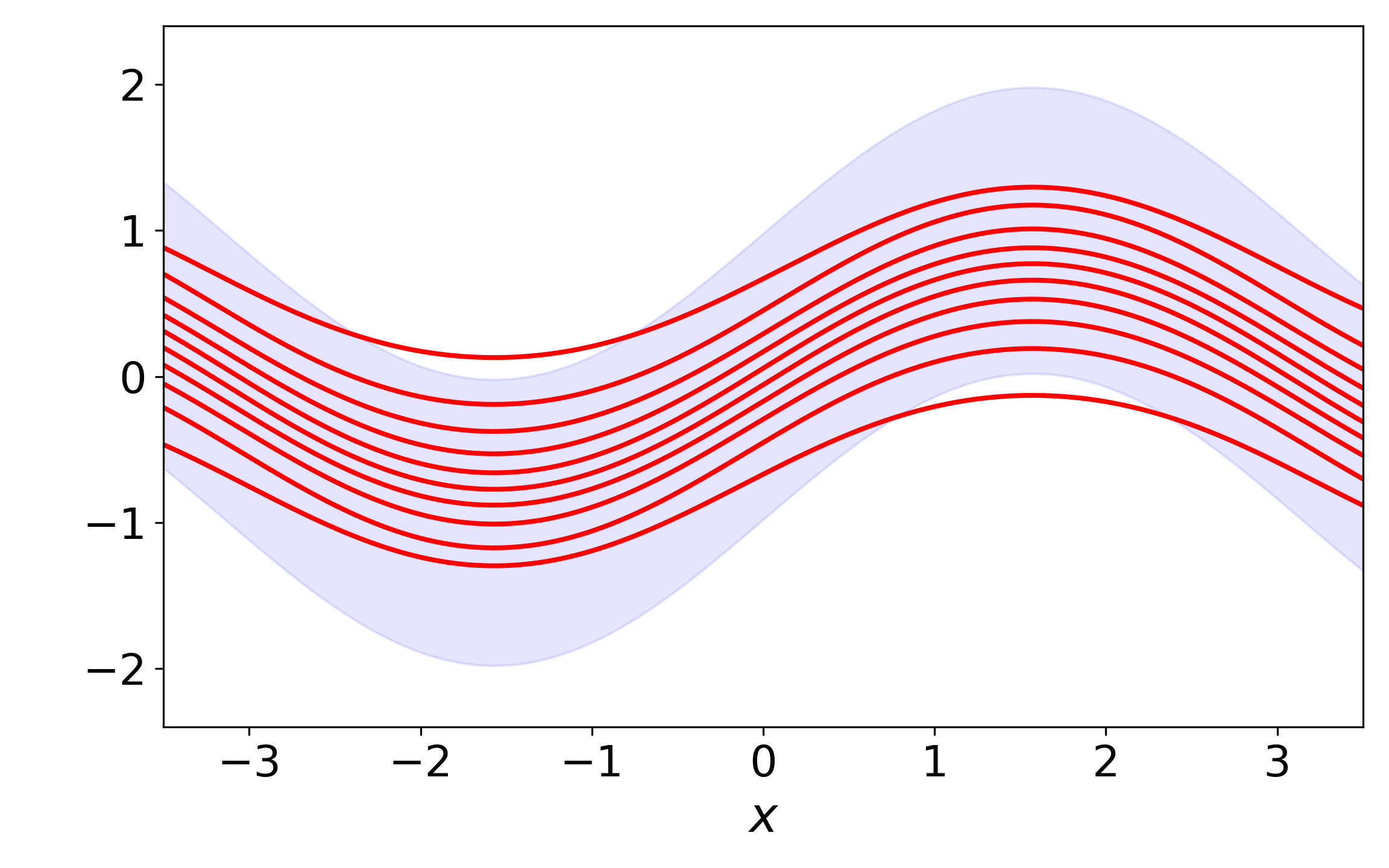}}
		\hfill
		\subcaptionbox{$100$ weak learners trained}{\includegraphics[width=0.315\textwidth]{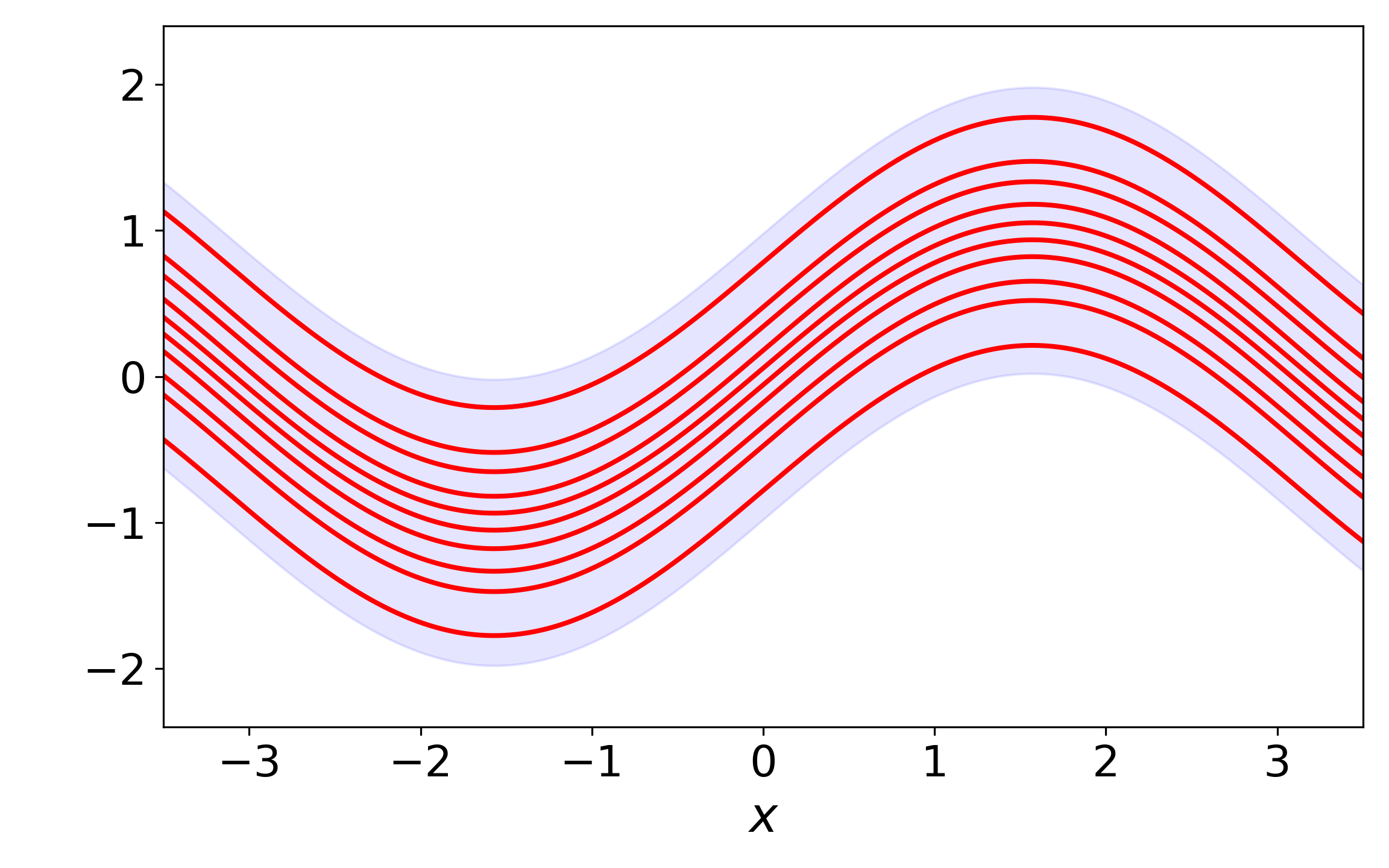}}
		\caption{Illustration of inputs and outputs of WGBoost trained on a training set $\{ x_i, \mu_i \}_{i=1}^{10}$ whose inputs are 10 grid points in $[-3.5, 3.5]$ and output distributions are each a normal distribution $\mu_i(\theta) = \mathcal{N}(\rv \mid \sin(x_i), 0.5)$ over $\theta \in \R$. The blue area indicates the $95$\% high probability region of the output distribution for each point. WGBoost returns $N$ particles (red lines) that predicts the output distribution for each input, where this illustration selects $N = 10$ and uses a Gaussian kernel regressor as each weaker learner of WGBoost.}\label{fig:illustration}
	\end{figure}

\begin{figure}[t]
		\centering
		\subcaptionbox{Bayesian learning of a model $f(x, w)$}{\includegraphics[width=0.475\textwidth]{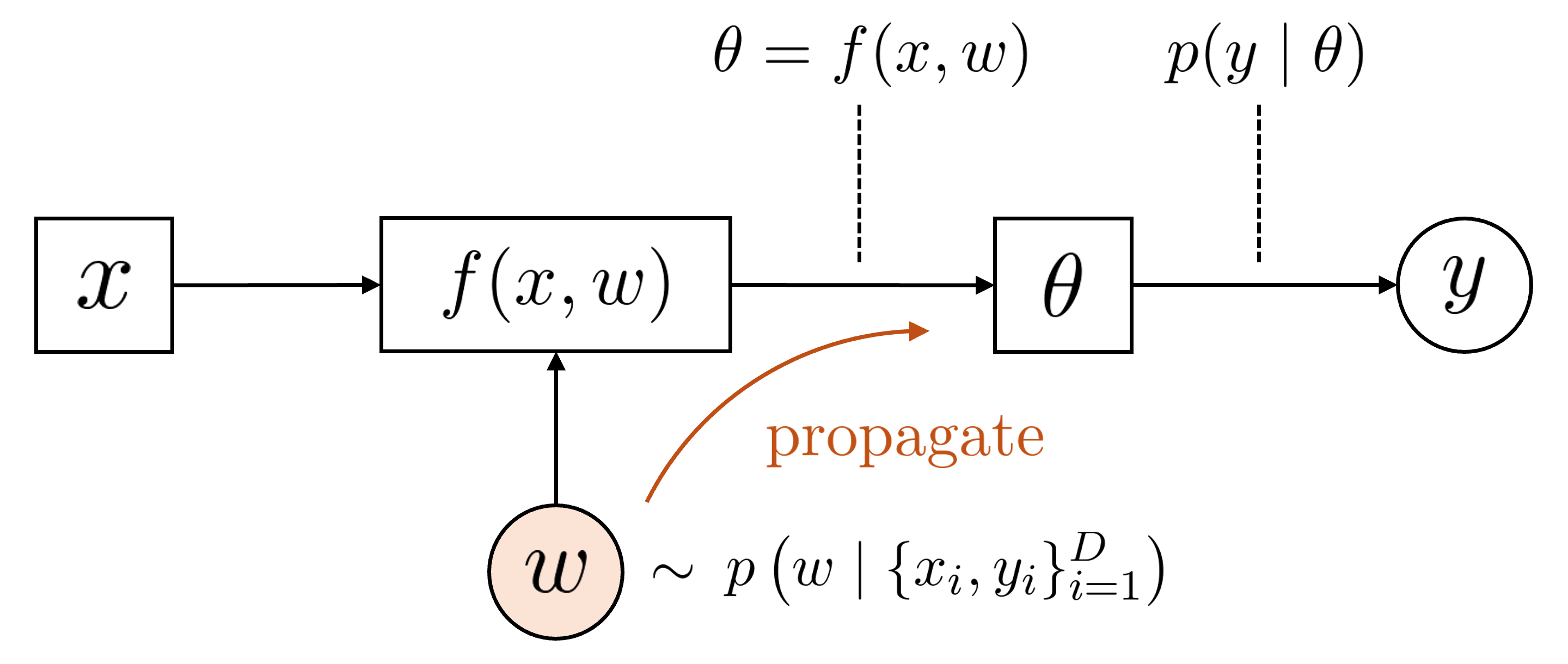}}
		\hfill
		\subcaptionbox{Evidential learning based on WGBoost}{\includegraphics[width=0.475\textwidth]{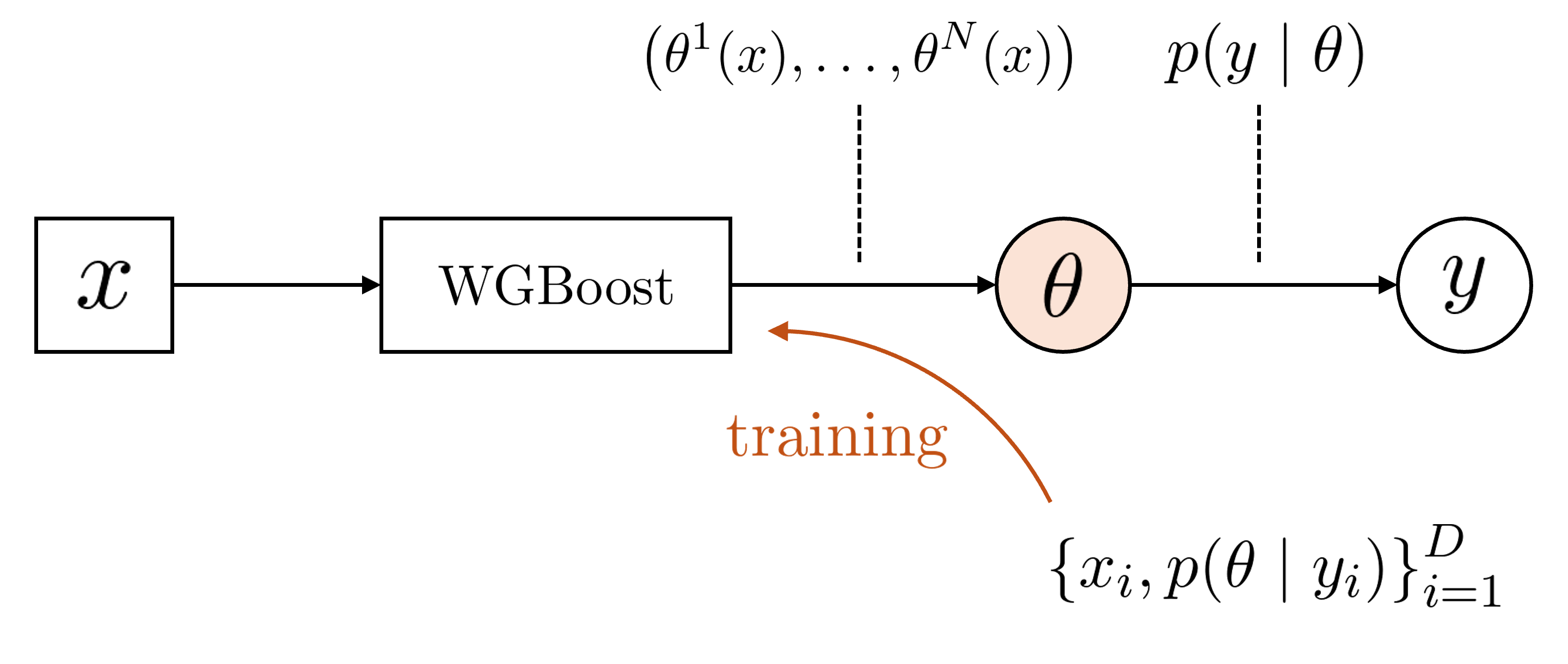}}
		\hfill
		\caption{Comparison of the pipeline of (a) Bayesian learning and (b) evidential learning based on WGBoost. The former uses the (global) posterior $p(w \mid \{ x_i, y_i \}_{i=1}^{D})$ of the model parameter $w$ conditional on all data, and samples multiple models from it. The latter uses the individual-level posterior $p(\theta \mid y_i)$ of the response parameter $\theta$ conditional on each individual datum $y_i$ as the output distribution in the training set, and trains WGBoost to directly returns a particle-based distributional estimate $p(\theta \mid x)$ of $\theta$ for each input $x$.}\label{fig:contrast}
	\end{figure}

	Motivated by this question, we formulate a general framework of Wasserstein gradient boosting (WGBoost). 
	WGBoost receives an input and returns a particle-based prediction of the output probability distribution.
	\Cref{fig:illustration} illustrates inputs and outputs of WGBoost.
	This paper considers application of WGBoost to evidential learning, where the individual-level posterior $p(\rv \mid y_i )$ of the response distribution $p(y \mid \rv)$ is used as the output distribution $\mu_i$ for each input $x_i$ of the training set.
	\Cref{fig:contrast} compares the pipeline of evidential learning with Bayesian learning.
	To the author's knowledge, WGBoost is the first framework that enables evidential learning (i) for boosted tree models and (ii) without closed form of individual-level posteriors.

	\textbf{Contributions }
	Our contributions are summarised as follows:
	
	\begin{itemize}
		\item \Cref{sec:section2} establishes the general framework of WGBoost. 
		It is a novel family of gradient boosting that returns a set of particles that approximates an output distribution assigned at each input. 
		In contrast to standard gradient boosting that fits a weak learner to the gradient of a loss function, WGBoost fits a weak learner to the estimated Wasserstein gradient of a loss functional over probability distributions.
		\item \Cref{sec:section3} establishes tree-based evidential learning based on WGBoost, with the loss functional specified by the Kullback–Leibler (KL) divergence.
		Following modern gradient-boosting libraries \cite{Chen2016,Ke2017} that use second-order gradient boosting (c.f.~\Cref{sec:section22}), we use a second-order WGBoost algorithm built on an approximate Wasserstein gradient and Hessian of the KL divergence (c.f.~\Cref{sec:section33,sec:section34}).
		\item 	\Cref{sec:section4} demonstrates the performance of probabilistic regression and classification with OOD detection on real-world tabular datasets in comparison with common uncertainty quantification methods.
	\end{itemize}
	
	\section{General Formulation of Wasserstein Gradient Boosting} \label{sec:section2}
	
	This section presents the general formulation of WGBoost.
	\Cref{sec:section21} recaps the notion of Wasserstein gradient flows, a `gradient' system of probability distributions that minimises an objective functional in the Wasserstein space.
	\Cref{sec:section22} recaps the notion of gradient boosting, a sequential ensemble method that fits a new weak learner to the `gradient' of the remaining loss at each iteration.
	\Cref{sec:section23} combines the above two notions to establish WGBoost, a novel family of gradient boosting whose output is a set of particles that approximates an output distribution assigned at each input.

	\noindent
	\textbf{Notation and Setting~}
	Let $\X$ and $\Y$ denote the space of inputs and responses in classification and regression.
	Suppose $\Theta = \R^d$.
	Let $\P2$ be the 2-Wasserstein space, that is, a set of all probability distributions on $\Theta$ with finite second moment equipped with the Wasserstein metric \cite{Villani2003}.
	We identify a probability distribution in $\P2$ with its density whenever it exits.
	Denote by $\odot$ and $\oslash$, respectively, elementwise multiplication and division of two vectors in $\R^d$.
	Let $\nabla$ be the gradient operator.
	Let $\nabla_{\text{d}}^2$ be a second-order gradient operator that takes the second derivative at each coordinate i.e.~$\nabla_{\text{d}}^2 f(\rv) = [ \partial^2 f(\rv) / \partial \rv_1^2 , \dots, \partial^2 f(\rv) / \partial \rv_d^2 ]^{\operatorname{T}} \in \R^d$.

	\subsection{Wasserstein Gradient Flow} \label{sec:section21}
	
	In the Euclidean space, a gradient flow of a function $f$ means a curve of points $x_t$ that solves a differential equation $( d / dt ) x_t = - \nabla f(x_t)$ from an initial value $x_0$.
	That is the continuous-time limit of gradient descent, which minimises the function $f$ as $t \to \infty$.
	A Wasserstein gradient flow means a curve of probability distributions $\mu_t$ minimising a given functional $\mathcal{F}$ on the 2-Wasserstein space $\P2$.
	The Wasserstein gradient flow $\mu_t$ is characterised as a solution of a partial differential equation, known as the \emph{continuity equation}:
	\begin{align}
		\frac{d}{d t} \mu_t = - \nabla \cdot \left( \mu_t \nabla_{W} \F(\mu_t) \right) \quad \text{given} \quad \mu_0 \in \P2 , \label{eq:continuity_equation}
	\end{align}
	where $\nabla_{W} \F(\mu): \Theta \to \Theta$ denotes the \emph{Wasserstein gradient} of $\F$ at $\mu$ \cite{Ambrosio2005,Santambrogio2017}.
	\Cref{apx:appendix_a} recaps the derivation of the Wasserstein gradient and presents the examples for several functionals.
	
	One of the elegant properties of the Wasserstein gradient flow is casting the infinite-dimensional optimisation of the functional $\F$ as a finite-dimensional particle update \cite{Villani2003}.
	The continuity equation \eqref{eq:continuity_equation} can be reformulated as a dynamical system of a random variable $\rv_t \sim \mu_t$, such that
	\begin{align}
		\frac{d}{d t} \rv_t = - \left[ \nabla_{W} \F(\mu_t) \right](\rv_t) \quad \text{given} \quad \rv_0 \sim \mu_0 , \label{eq:wg_update}
	\end{align} 
	in the sense that the law $\mu_t$ of the random variable $\rv_t$ is a weak solution of the continuity equation.
	Consider the case where the initial measure $\mu_0$ is set to the empirical distribution $\hat{\mu}_0$ of $N$ particles $\{ \rv_0^n \}_{n=1}^{N}$.
	Discretising the continuous-time system \eqref{eq:wg_update} by the Euler method with a small step size $\nu > 0$ yields an iterative update scheme of $N$ particles $\{ \rv_m^n \}_{n=1}^{N}$ from step $m = 0$:
	\begin{align}
		\begin{bmatrix}
			\rv_{m+1}^1 \\
			\vdots \\
			\rv_{m+1}^N
		\end{bmatrix} 
		= 
		\begin{bmatrix}
			\rv_{m}^1 \\
			\vdots \\
			\rv_{m}^N
		\end{bmatrix}
		+ \nu
		\begin{bmatrix}
			- [ \nabla_{W} \F(\hat{\mu}_m) ]( \rv_{m}^1 ) \\
			\vdots \\
			- [ \nabla_{W} \F(\hat{\mu}_m) ]( \rv_{m}^N )
		\end{bmatrix} , \label{eq:wg_particle_update}
	\end{align}
	where $\hat{\mu}_m$ denotes the empirical distribution of the particles $\{ \rv_m^n \}_{n=1}^{N}$ at step $m$.
	
	In practice, it is common that the Wasserstein gradient of a chosen functional $\F$ is not well-defined for empirical distributions.
	In such cases, the particle update scheme \eqref{eq:wg_particle_update} is not directly applicable because it depends on the Wasserstein gradient $\nabla_{W} \F(\hat{\mu}_m)$ at the empirical distribution $\hat{\mu}_m$.
	For example, the KL divergence $\F(\mu) = \KL(\mu \mid \pi)$ with a reference distribution $\pi$ has such a Wasserstein gradient $[ \nabla_{W} \F(\mu) ](\rv) = - ( \nabla \log \pi(\rv) - \nabla \log \mu(\rv) )$ ill-defined when $\mu$ is an empirical distribution.
	Hence, one often uses an estimate or approximation of the Wasserstein gradient of a functional $\F$ that is well-defined for empirical distributions, in order to perform the particle update scheme \eqref{eq:wg_particle_update} approximately \cite[e.g.][]{Liu2017,Carrillo2019,Wang2022,Maoutsa2020,He20202}.
	Our application of WGBoost in \Cref{sec:section3} uses the `smoothed' Wasserstein gradient of the KL divergence \cite{Liu2017} recapped later.

	\subsection{Gradient Boosting} \label{sec:section22}
	
	Gradient boosting \cite{Friedman2001} is a sequential ensemble method of $M$ multiple weak learners $f_1, \dots, f_M$.
	It iteratively constructs an ensemble $\B_m$ of $m$ weak learners $f_1, \dots, f_m$ from step $m = 0$ to $M$.
	Given the current ensemble $\B_m$ at step $m$, it trains a new weak learner $f_{m+1}$ and constructs the next ensemble $\B_{m+1}$ by
	\begin{align}
		\B_{m+1}(x) = \B_m(x) + \nu f_{m+1}(x) \label{eq:gb_ensemble}
	\end{align}
	where $\nu$ is a shrinkage hyperparameter called a \emph{learning rate}.
	The initial state of the ensemble $\B_0(x)$ at step $m = 0$ is typically set to a constant that best fits the data.
	Any learning algorithm can be used as a weak learner in principle, although tree-based algorithms are most used \cite{Duan2020}.
	
	The fundamental idea of gradient boosting is to train the new weak learner $f_{m+1}$ to approximate the negative gradient of the remaining error of the current ensemble $F_m$.
	Suppose that outputs are vectors in $\R^d$ and that a loss function $L$ measures the remaining error at each data point $R_i(\B_m(x_i)) := L( \B_m(x_i), y_i )$.
	The new weak learner $f_{m+1}$ is fitted to the set $\{ x_i, g_i \}_{i=1}^{D}$, where the target variable $g_i$ is specified as
	\begin{align}
		g_i = - \nabla R_i( \B_m(x_i) ) \in \R^d . \label{eq:gb_gradient}
	\end{align}
	The target variable $g_i$ is often called a \emph{pseudo residual}.
	At every fixed input $x_i$, the boosting scheme \eqref{eq:gb_ensemble} updates the output of the current ensemble $\B_m(x_i)$ in the steepest descent direction of the error $R_i(\B_m(x_i))$.
	Although \cite{Friedman2001} originally suggested performing an additional line search to determine a scaling constant for each weak learner, the line search has been reported to have a negligible influence on performance \cite{Buhlmann2007}.
	
	In modern gradient-boosting libraries such as XGBoost \cite{Chen2016} and LightGBM \cite{Ke2017}, the standard practice is to use the diagonal (coordinatewise) Newton direction of the remaining error for the target variable of the new weak learner $f_{m+1}$.
	In this case, the new base leaner $f_{m+1}$ is fitted to the set $\{ x_i, g_i \oslash h_i \}_{i=1}^{n}$, where the negative gradient $g_i$ is divided elementwise by the Hessian diagonal $h_i$ specified as
	\begin{align}
		h_i = \nabla_{\text{d}}^2 R_i( \B_m(x_i) ) \in \R^{d} .
	\end{align}
	The target variable $g_i \oslash h_i$ is the diagonal Newton direction that minimises the second-order Taylor approximation of the remaining error for each coordinate independently.
	Combining second-order gradient boosting with tree-based weak learners has demonstrated exceptional scalability and performance \cite{Grinsztajn2022,Florek2023}.
	Although it is possible to use the `full' Newton direction as the target variable of each weak learner, the impracticality of the full Newton direction has been pointed out \cite[e.g.][]{Zhang2021,Chen2015}.
	In addition, the coordinatewise computability of the diagonal Newton direction is suitable for popular gradient-boosting tree algorithms \cite{Zhang2021}.

	\subsection{Wasserstein Gradient Boosting} \label{sec:section23}
	
	Now we consider the setting of `distribution-valued' supervised learning, where we are given a training set of input vectors and output distributions $\{ x_i, \mu_i \}_{i=1}^{D} \subset \X \times \P2$.
	Our goal is to construct a model that receives an input and returns a set of $N$ particles whose empirical distribution approximates the output distribution.
	We specify a loss functional $\D(\cdot \mid \cdot)$ between two probability distributions---such as the KL divergence---to measure the remaining error $\F_i(\cdot) = \D(\cdot \mid \mu_i)$ of the model output for each $i$-th training output distribution $\mu_i$.
	Our idea is to combine gradient boosting with Wasserstein gradient, where we iteratively construct a set of $N$ boosting ensembles $\B_m^1, \dots, \B_m^N$---each of which consists of $m$ weak learners---from step $m = 0$ to $M$.
	
	Here, the output $\B_m^n(x)$ of each $n$-th boosting ensemble represents the $n$-th output particle for an input $x$.
	Given the current set of $N$ ensembles $\B_m^1, \dots, \B_m^N$ at step $m$, WGBoost trains a set of $N$ new weak learners $f_{m+1}^1, \dots, f_{m+1}^N$ and computes the next set of $N$ ensembles $\B_{m+1}^1, \dots, \B_{m+1}^N$ by
	\begin{align}
		\begin{bmatrix}
			\B_{m+1}^1(x) \\
			\vdots \\
			\B_{m+1}^N(x)
		\end{bmatrix}
		= 
		\begin{bmatrix}
			\B_m^1(x) \\
			\vdots \\
			\B_m^N(x)
		\end{bmatrix}
		+ \nu
		\begin{bmatrix}
			f_{m+1}^1(x) \\
			\vdots \\
			f_{m+1}^N(x)
		\end{bmatrix} \label{eq:wgb_update}
	\end{align}
	where $\nu$ is a learning rate.
	Similarly to standard gradient boosting, we set the initial state of $N$ ensembles $\B_0^1, \dots, \B_0^N$ at step $m = 0$ to a set of given constants.
	Throughout, denote by $\A_{m, i}$ the empirical distribution of the $N$ output particles $\B_m^1(x_i), \dots, \B_m^N(x_i)$ for each $i$-th training input $x_i$.
	
	As discussed in \Cref{sec:section21}, the Wasserstein gradient often needs to be estimated when a distribution is an empirical distribution.
	For better presentation, let $\G_i(\mu)$ denote an estimate of the Wasserstein gradient of the $i$-th remaining error $\F_i(\mu)$ at arbitrary $\mu$.
	If the original Wasserstein gradient is well-defined for all $\mu$, it is a trivial estimate to use as $\G_i(\mu)$.
	Otherwise, any suitable estimate can be used as $\G_i(\mu)$.
	The foundamental idea of WGBoost is to train the $n$-th new learner $f_{m+1}^n$ to approximate the estimated Wasserstein gradient $- \G_i(\A_{m, i})$ evaluated at the $n$-th boosting output $\B_m^n(x_i)$ for each $x_i$, so that,
	\begin{align}
		\begin{bmatrix}
			f_{m+1}^1(x_i) \\
			\vdots \\
			f_{m+1}^N(x_i)
		\end{bmatrix}
		\approx
		\begin{bmatrix}
			- \left[ \G_i\left( \A_{m, i} \right) \right]\left( \B_m^1(x_i) \right) \\
			\vdots \\
			- \left[ \G_i\left( \A_{m, i} \right) \right]\left( \B_m^N(x_i) \right)
		\end{bmatrix} . \label{eq:wgb_training}
	\end{align}
	At every fixed $x_i$, the boosting scheme \eqref{eq:wgb_update} approximates the particle update scheme \eqref{eq:wg_particle_update} for the output particles $\B_m^1(x_i), \dots, \B_m^N(x_i)$ under the estimated Wasserstein gradint $\nabla_W \F_i(\A_{m, i})$, by which each boosting output is updated in the direction to decrease the remianing error $\F_i(\A_{m, i}) = \D(\A_{m, i} \mid \mu_i)$ at step $m$.
	
	The general procedure of WGBoost is summarised in Algorithm \ref{alg:wgb}.
	For our application, we focus on the KL divergence as a choice of the loss functional $\D(\cdot \mid \cdot)$ and use the Wasserstein gradient estimate based on kernel smoothing in \Cref{sec:section3}.
	\Cref{apx:appendix_a} presents examples of Wasserstein gradients of several divergences.
	\Cref{fig:illustration} illustrates the output of WGBoost using a toy output distribution $\mu_i(\cdot) = \mathcal{N}(\cdot \mid \sin(x_i), 0.5)$.
	
	\begin{algorithm}[t]
		\caption{Wasserstein Gradient Boosting} \label{alg:wgb}
		\KwIn{training set $\{ x_i, \mu_i \}_{i=1}^{D}$ of input $x_i \in \X$ and output distribution $\mu_i \in \P2$}
		\Parameter{loss functional $\D(\cdot \mid \cdot)$, estimate $\G_i(\mu)$ of the Wasserstein gradient of $\D(\mu \mid \mu_i)$, particle number $N$, iteration $M$, learning rate $\nu$, weak learner $f$, initial constants $( \vartheta_0^1, \dots, \vartheta_0^N )$}
		\KwOut{set of $N$ boosting ensembles $( \B_M^1, \dots, \B_M^N )$ at final step $M$}
		$( \B_0^1(\cdot), \dots, \B_0^N(\cdot) ) \gets ( \vartheta_0^1, \dots, \vartheta_0^N )$ \hfill $\rhd$ set initial state of $N$ boosting ensembles\\
		\For{$m \gets 0, \dots, M-1$}{
			\For{$i \gets 1, \dots, D$}{
				$\hat{\mu}_{m,i} \gets \text{empirical distribution of set of $N$ output values } (\B_m^1(x_i), \dots, \B_m^N(x_i))$\text{ for input $x_i$}\\
				\For{$n \gets 1, \dots, N$}{
					$g_i^n \gets - \left[ \G_i(\A_{m, i}) \right]\left( \B_m^n(x_i) \right)$ \hfill $\rhd$ Wasserstein gradient evaluated at $n$-th output value for $x_i$
				}
			}
			\For{$n \gets 1, \dots, N$}{
				$f_{m+1}^n \gets \text{fit}\left( \left\{ x_i, g_i^n \right\}_{i=1}^{D} \right)$ \hfill $\rhd$ fit $n$-th new weak learner to Wasserstein gradients\\
				$\B_{m+1}^n(\cdot) \gets \B_m^n(\cdot) + \nu f_{m+1}^n(\cdot)$ \hfill $\rhd$ get next state of $n$-th boosting ensemble
			}
		}
	\end{algorithm}

	\begin{remark}[\textbf{Stochastic WGBoost}] \label{rm:stochastic_WGB}
		Stochastic gradient boosting \cite{Friedman2002} uses only a randomly sampled subset of data to fit a new weak learner at each step $m$ to reduce the computational cost.
		The same subsampling approach can be applied for WGBoost whenever the dataset is large.
	\end{remark}
		
	\begin{remark}[\textbf{Second-Order WGBoost}] \label{rm:second_order_WGB}
		If an estimate of the Wasserstein `Hessian' of the remaining error $\F_i$ is available, the Newton direction of $\F_i$ may also be computable \cite[e.g.][]{Detommaso2018,Wang2020}.
		We can immediately implement a second-order WGBoost algorithm by plugging such a Newton direction into $\G_i( \mu )$ in Algorithm \ref{alg:wgb}.
		Our default WGBoost algorithm for tree-based evidential learning is built on a diagonal approximate Newton direction of the KL divergence, aligning with the standard practice in modern gradient-boosting libraries to use the diagonal Newton direction.
	\end{remark}

	\section{Default Setting for Tree-Based Evidential Learning} \label{sec:section3}
	
	This section provides the default setting to implement a concrete WGBoost algorithm for evidential learning, which enables classification and regression with predictive uncertainty.
	The individual-level posterior $p(\rv \mid y_i)$ of a response distribution $p(y \mid \rv)$ is used as the output distribution $\mu_i$ in the training set $\{ x_i, \mu_i \}_{i=1}^{D}$ of WGBoost.
	\Cref{sec:section31} recaps derivation of the individual-level posterior $p(\rv \mid y_i)$, followed by the default choice of the prior discussed in \Cref{sec:section32}.
	We choose the KL divergence as a loss functional $\D(\cdot \mid \cdot)$ of WGBoost.
	\Cref{sec:section33} recaps a widely-used approximation of the Wasserstein gradient of the KL divergence based on kernel smoothing \cite{Liu2017}.
	A further advantage of the kernel smoothing approach is that the approximate Wasserstein Hessian is available, with which \Cref{sec:section34} establishes a second-order WGBoost algorithm similarly to modern gradient-boosting libraries.

	\subsection{Individual-Level Posteriors as Output Distributions} \label{sec:section31}
	
	Suppose that a response distribution $p(y \mid \rv)$ is specified for a response variable $y$ of one's classification or regression problem, as is typically done.
	Suppose further that a prior distribution $p_i(\rv)$ of the response parameter $\rv$ is specified at each individual observed input $x_i$.
	At each individual observation $(x_i, y_i)$, the response-distribution likelihood $p(y_i \mid \rv)$ and the prior $p_i(\rv)$ determine the individual-level posterior
	\begin{align}
		p(\rv \mid y_i) \propto p(y_i \mid \rv) p_i(\rv) . \label{eq:bayes_posterior}
	\end{align}
	The individual-level posterior is used as the output distribution $\mu_i$ for each input $x_i$ in the training set $\{ x_i, \mu_i \}_{i=1}^{D}$ of WGBoost.
	We apply the framework of WGBoost to construct a model that returns a set of particles that approximates the output distribution $\mu_i(\cdot) = p(\cdot \mid y_i)$ for each observed input $x_i$. 
	
	For a new input $x$, the trained WGBoost model returns a set of particles $( \rv^1(x), \dots, \rv^N(x) )$ that is a nonparametric distributional estimate $p(\rv \mid x)$ of the response parameter.
	Based on the output particles, a predictive distribution $p(y \mid x)$ of the response $y$ for each new input $x$ can be defined via marginalisation:
	\begin{align}
		p(y \mid x) = \int_\Theta p(y \mid \rv) p(\rv \mid x) d \rv = \frac{1}{N} \sum_{i=1}^{N} p\left( y \mid \rv^i(x) \right) . \label{eq:BMA}
	\end{align}
	A point prediction $\hat{y}$ for each new input $x$ can also be defined via the `individual-level' Bayes action:
	\begin{align}
		\hat{y} = \text{argmin}_{y \in \Y} ~ \int_\Theta U(y, \rv) p(\rv \mid x) d \rv = \text{argmin}_{y \in \Y} ~ \frac{1}{N} \sum_{i=1}^{N} U(y, \rv^i(x))  , \label{eq:BA}
	\end{align}
	which is the minimiser of the average risk of a given utility $U: \Y \times \Theta \to \R$.
	For example, if the utility is a quadratic function $U(y, \rv) = (y - \rv)^2$, the Bayes action is the mean of the output particles $( \rv^1(x), \dots, \rv^N(x) )$.

	In general, the explicit form of the individual-level posterior $p(\rv \mid y_i)$ is known only up to the normalising constant.
	The WGBoost algorithm for evidential learning, provided in \Cref{sec:section34}, requires no normalising constant of the individual-level posterior $p(\rv \mid y_i)$.
	It depends only on the log-gradient of the individual-level posterior $\nabla \log p(\rv \mid y_i) = \nabla p(\rv \mid y_i) / p(\rv \mid y_i)$, cancelling the normalising constant by fraction. 
	Hence, knowing the form of the response-distribution likelihood $p(y_i \mid \rv)$ and the prior $p_i(\rv)$ suffices.
	
	\begin{remark}[\textbf{Difference from Bayesian Learning}]
		Bayesian learning of a given model $f(x, w)$ uses the posterior $p(w \mid \{ x_i , y_i \}_{i=1}^{D})$ of the model parameter $w$ conditional on all data.
		The predictive distribution $p(y \mid x)$ of the response $y$ in Bayesian learning is defined via marginalisation of the model parameter $w$, that is, $p(y \mid x) = \int_\Theta p(y \mid \rv = f(x, w)) p(w \mid \{ x_i , y_i \}_{i=1}^{D}) d w$.
		In contrast, WGBoost learns the nonparametric distributional estimate $p(\rv \mid x)$ directly from the training set $\{ x_i, \mu_i \}_{i=1}^{D}$, involving no marginalisation of the model parameter $w$ that is often exceedingly high dimensional in machine learning.
	\end{remark}

	\subsection{Choice of Individual-Level Priors} \label{sec:section32}
	
	The prior $p_i(\rv)$ of the response parameter $\rv$ is specified at each individual observed input $x_i$.
	The approach to specifying the prior may differ depending on whether past data are available.
	When past data are available, past data can be utilised in any possible way to elicit a reasonable prior for future data.
	When no past data are available, we recommend the use of a noninformative prior that have been developed as a sensible choice of prior in the absence of past data; see \cite[e.g.][]{Ghosh2011} for the introduction.
	To avoid numerical errors, if a noninformative prior is improper (nonintegrable) as is often the case, we recommend the use of a proper probability distribution that approximates the noninformative prior sufficiently well.

	\begin{example}[\textbf{Normal Location-Scale}] \label{ex:normal}
		Consider regression with a scalar-valued response variable $y \in \R$.
		A normal location-scale distribution $\mathcal{N}(y \mid m, \sigma)$ has the mean and scale parameters $m \in \R$ and $\sigma \in (0, \infty)$.
		A typical noninformative prior of $m$ and $\sigma$ are given by $1$ and $1 / \sigma$ respectively, which are improper.
		At every observation $(x_i, y_i)$, we use a normal prior $\mathcal{N}(m \mid 0, \sigma_0)$ over $m$ and an inverse gamma prior $\operatorname{IG}(\sigma \mid \alpha_0, \beta_0)$ over $\sigma$, with the hyperparameters $\sigma_0 = 10$ and $\alpha_0 = \beta_0 = 0.01$, which approximate the non-informative priors.
	\end{example}
	
	\begin{example}[\textbf{Categorical}] \label{ex:categorical}
		Consider classification with a $k$-class label response variable $y \in \{ 1, \dots, k \}$.
		A categorical distribution $\mathcal{C}(y \mid q)$ has a class probability parameter $q = ( q_1, \dots, q_k )$ in the $k$-dimensional simplex $\Delta_k$.
		If $k = 2$, it corresponds to the Bernoulli distribution.
		A typical noninformative prior of $q$ is given by $1 / \prod_{i=1}^{k} q_i$.
		At every observation $(x_i, y_i)$, we use the logistic normal prior---a multivariate generalisation of the logit normal distribution \cite{Aitchison1980}---over $q$ with the mean $0$ and identity covariance matrix scaled by $10$.
	\end{example}
	
	In \Cref{sec:section2}, we have supposed that $\Theta = \R^d$ for some dimension $d$. 
	Any parameter that lies in a subset of the Euclidean space (e.g. $\sigma$) can be reparametrised as one in the Euclidean space (e.g. $\log \sigma$).
	\Cref{apx:appendix_e} details the reparametrisation used for the experiment.
	If a dataset has scalar outputs and they have a low or high order of magnitude, we also recommend standardising the outputs to adjust the magnitude.

	\subsection{Approximate Wasserstein Gradient of KL Divergence} \label{sec:section33}

	The loss functional $\D(\mu \mid \mu_i)$ considered in this setting is the KL divergence $\KL(\mu \mid \mu_i)$.
	A computational challenge of the KL divergence is that the associated Wasserstein gradient $\left[ \G_i^{\text{KL}}(\mu) \right]( \rv ) := - \left( \nabla \log \pi_i(\rv) - \nabla \log \mu(\rv) \right)$ is not well-defined for empirical distributions.
	A particularly successful approach to finding a well-defined approximation of the Wasserstein gradient---which originates in \cite{Liu2016} and has been applied in wide contexts \cite{Liu2017,Wang2018,Lambert2021}---is to smooth the original Wasserstein gradient through a kernel integral operator $\int_\Theta [ \G_i^{\text{KL}}(\mu) ](\rv^*) k(\rv, \rv^*) d \mu(\rv^*)$ \cite{Korba2020}.
	By integration-by-part (see \cite[e.g.][]{Liu2016}), the smoothed Wasserstein gradient---denoted $\G^*_i(\mu)$---falls into the following form that is well-defined for any distribution $\mu$:
	\begin{align}
		\left[ \G^*_i(\mu) \right]( \rv ) & := - \E_{\rv^* \sim \mu}\Big[ \nabla \log \mu_i(\rv^*) k(\rv, \rv^*) + \nabla k(\rv, \rv^*) \Big] \in \R^d , \label{eq:swg}
	\end{align}
	where $\nabla k(\rv, \rv^*)$ denotes the gradient of $k$ with respect to the first argument $\rv$.
	An approximate Wasserstein gradient flow based on the smoothed Wasserstein gradient $\G^*_i(\mu)$ is called the Stein variational gradient descent \cite{Liu2016} or kernelised Wasserstein gradient flow \cite{Chewi2020}.
	In most cases, the kernel $k$ is set to the Gaussian kernel $k(\rv, \rv^*) = \exp( - \| \rv - \rv^* \|^2 / h )$ with the scale hyperparameter $h > 0$.
	We use the Gaussian kernel with the scale hyperparameter $h = 0.1$ throughout this work.
	
	Another common approach to approximating the Wasserstein gradient flow of the KL divergence is the Langevin diffusion approach \cite{Wibisono2018}.
	The discretised algorithm, called the unadjusted Langevin algorithm \cite{Roberts1996}, is a stochastic particle update scheme that adds a Gaussian noise at every iteration.
	However, several known challenges, such as asymptotic bias and slow convergence, often necessitate an ad-hoc adjustment of the algorithm \cite{Wibisono2018}.
 	\Cref{apx:appendix_b} discusses a variant of WGBoost built on the Langevin algorithm, although it is not considered the default implementation.

	\subsection{Second-Order Implementation of WGBoost} \label{sec:section34}
	
	Following the standard practice in modern gradient-boosting libraries \cite{Chen2016,Ke2017} to use the diagonal Newton direction, we further consider a diagonal (coordinatewise) approximate Wasserstein Newton direction of the KL divergence.
	In a similar manner to the smoothed Wasserstein gradient \eqref{eq:swg}, the approximate Wasserstein Hessian of each KL divergence $\KL(\mu \mid \mu_i)$ can be obtained by the kernel smoothing.
	The diagonal of the approximate Wasserstein Hessian, denoted $\H_i^*(\mu)$, is defined by
	\begin{align}
		\left[ \H_i^*(\mu) \right]( \rv ) := \E_{\rv^* \sim \mu}\Big[ - \nabla_{\text{d}}^2 \log \mu_i(\rv^*) k(\rv, \rv^*)^2 + \nabla k(\rv, \rv^*) \odot \nabla k(\rv, \rv^*) \Big] \in \R^d . \label{eq:sswg}
	\end{align}
	The diagonal approximate Wasserstein Newton direction of each KL divergence is then defined by $- \left[ \G_i^*(\mu) \right]( \cdot ) \oslash \left[ \H_i^*(\mu) \right]( \cdot )$.
	\Cref{apx:appendix_c} provides the derivation based on \cite{Detommaso2018} who derived the Newton direction of the KL divergence in the context of nonparametric variational inference.
	The second-order WGBoost algorithm is established by plugging it into $\G_i(\mu)$ in Algorithm \ref{alg:wgb} i.e.~setting
	\begin{align}
		\left[ \G_i(\mu) \right]( \cdot ) = \left[ \G_i^*(\mu) \right]( \cdot ) \oslash \left[ \H_i^*(\mu) \right]( \cdot ) . \label{eq:diagonal_newton_wgkl}
	\end{align}
	Algorithm \ref{alg:wgb} under the setting \eqref{eq:diagonal_newton_wgkl} is considered our default WGBoost algorithm for evidential learning.
	We refer this algorithm to as the \emph{Wasserstein-boosted evidential learning} (WEvidential). 
	The explicit pseudocode is provided in Algorithm \ref{alg:wel} for full clarity.
	
	\begin{algorithm}[t]
		\caption{Wasserstein-Boosted Evidential Learning} \label{alg:wel}
		\KwIn{dataset $\{ x_i, y_i \}_{i=1}^{D}$ of input $x_i$ and response $y_i$ of classification or regression}
		\Parameter{individual-level posterior $p(\rv \mid y_i)$ of response distribution $p(y \mid \rv)$ conditional on each $y_i$, particle number $N$, iteration $M$, learning rate $\nu$, weak learner $f$, initial constants $\{ \vartheta_0^n \}_{n=1}^{N}$}
		\KwOut{set of $N$ boosting ensembles $( \B_M^1, \dots, \B_M^N )$ at final step $M$}
		$( \B_0^1(\cdot), \dots, \B_0^N(\cdot) ) \gets ( \vartheta_0^1, \dots, \vartheta_0^N)$ \hfill $\rhd$ set initial state of $N$ boostings\\
		\For{$m \gets 0, \dots, M-1$}{
			\For{$i \gets 1, \dots, D$}{
				$\hat{\mu}_{m,i} \gets \text{empirical distribution of set of $N$ output values } (\B_m^1(x_i), \dots, \B_m^N(x_i))$\text{ for input $x_i$}\\
				\For{$n \gets 1, \dots, N$}{
					$g_i^n \gets \E_{\rv^* \sim \hat{\mu}_{m,i}}[ \nabla \log p(\rv^* | y_i) k(\B_m^n(x_i), \rv^*) + \nabla k(\B_m^n(x_i), \rv^*) ]$\\
					$h_i^n \gets \E_{\rv^* \sim \hat{\mu}_{m,i}}[ - \nabla_{\text{d}}^2 \log p(\rv^* | y_i) k(\B_m^n(x_i), \rv^*)^2 + \nabla k(\B_m^n(x_i), \rv^*) \odot \nabla k(\B_m^n(x_i), \rv^*) ]$\\
				}
			}
			\For{$n \gets 1, \dots, N$}{
				$f_{m+1}^n \gets \text{fit}\left( \left\{ x_i, g_i^n \oslash h_i^n \right\}_{i=1}^{D} \right)$ \hfill $\rhd$ fit $n$-th new tree regressor to approximate Newton directions\\
				$\B_{m+1}^n(\cdot) \gets \B_m^n(\cdot) + \nu f_{m+1}^n(\cdot)$ \hfill $\rhd$ set next state of $n$-th boosting
			}
		}
	\end{algorithm}
	
	\begin{remark}[\textbf{Computation}]
		The diagonal Newton direction has a clear computational benefit in that only elementwise division is involved.
		The computational complexity is the same as that for the smoothed Wasserstein gradient, scaling linearly to both the particle number $N$ and the particle dimension $d$.
		Hence, there is essentially \emph{no reason not to use} the diagonal Newton direction instead of the smoothed Wasserstein gradient.
		Although it is possible to use the full Newton direction with no diagonal approximation, the inverse and product of $( N \times d  ) \times (N \times d)$ matrices are required at every computation of the direction (c.f.~\Cref{apx:appendix_e}).
		\Cref{apx:appendix_e} presents a simulation study to compare computational time and convergence speed of WGBoost algorithms implemented with four different estimates of the Wasserstein gradient.
	\end{remark}

	\section{Applications with Real-world Tabular Data} \label{sec:section4}
	
	We empirically demonstrate the performance of the WGBoost algorithm through three applications using real-world tabular data.
	The first application illustrates the output of WGBoost through a simple conditional density estimation.
	The second application benchmarks the regression performance on nine real-world datasets \cite{Hernandez-Lobato2015}.
	The third application examines the classification and OOD detection performance on the real-world datasets used in \cite{Charpentier2020}.
	The source code is available in \url{https://github.com/takuomatsubara/WGBoost}.
	
	\textbf{Common Hyperparameters:}
	Throughout, we set the number of output particles $N$ to $10$ and set each weak learner $f$ to the decision tree regressor \cite{Breiman1984} with maximum depth $1$ for the first application and $3$ for the rest.
	We set the learning rate $\nu$ to $0.1$ for regression and $0.4$ for classification.
	\Cref{apx:appendix_f} contains further details, including a choice of the initial constant $\{ \vartheta_0^n \}_{n=1}^{N}$.

	\subsection{Illustrative Conditional Density Estimation} \label{sec:section41}
	
	This section illustrates the output of the WGBoost algorithm by estimating a conditional density $p(y \mid x)$ from one-dimensional scalar inputs and outputs $\{ x_i, y_i \}_{i=1}^{D}$.
	The normal output distribution $\mathcal{N}(y \mid m, \sigma)$ and the prior $p_i(m, \sigma)$ in \Cref{ex:normal} were used to define the individual-level posterior $p(m, \sigma \mid y_i)$, in which case the output of the WGBoost algorithm is a set of $10$ particles $\{ ( m^n(x), \sigma^n(x) ) \}_{n=1}^{10}$ of the mean and scale parameters for each input $x$.
	We set the number of weak learners $M$ to $500$.
	
\begin{figure}[h]
		\centering
		\hfill
		\includegraphics[width=0.45\textwidth]{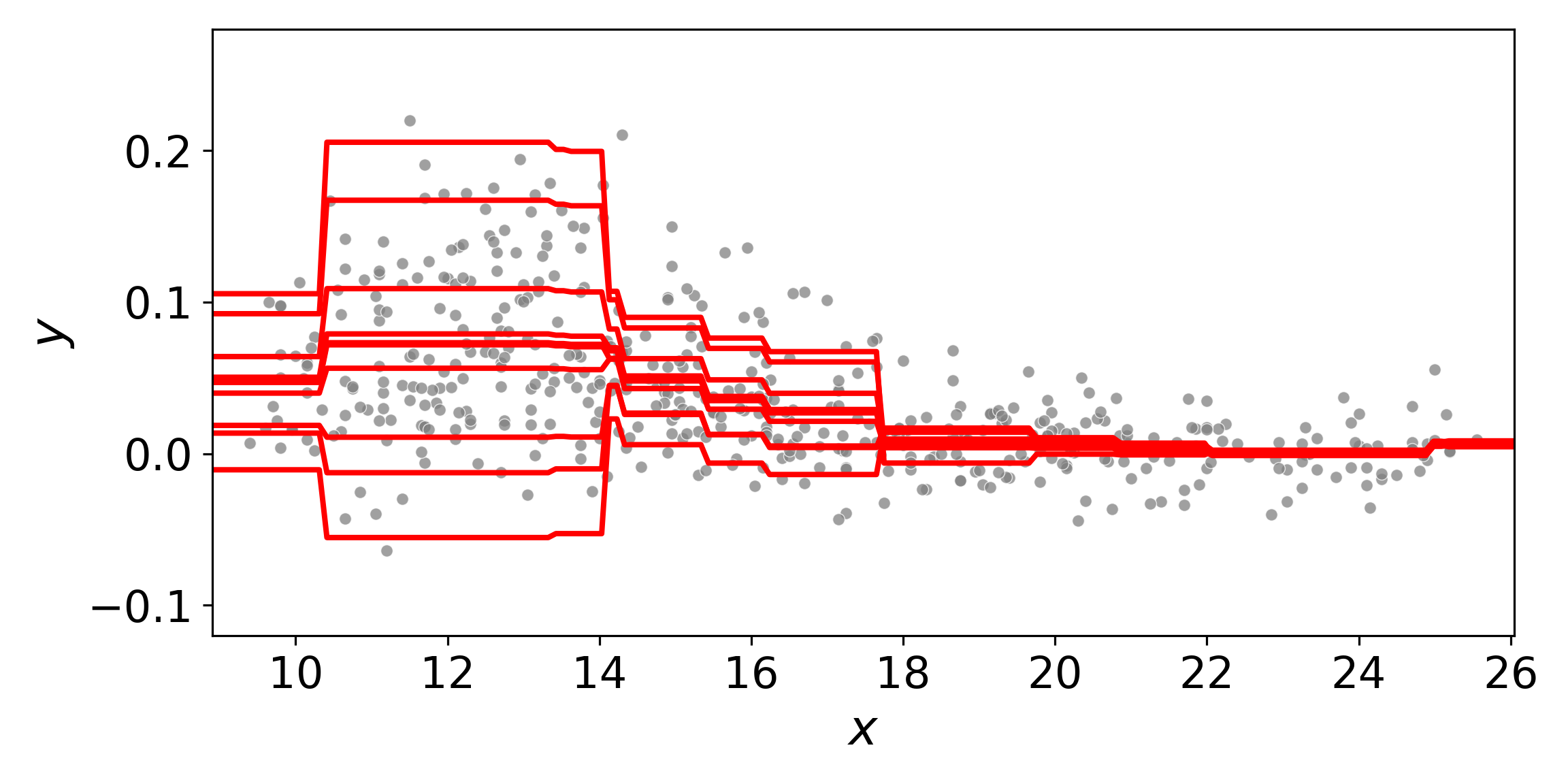}
		\hfill
		\includegraphics[width=0.45\textwidth]{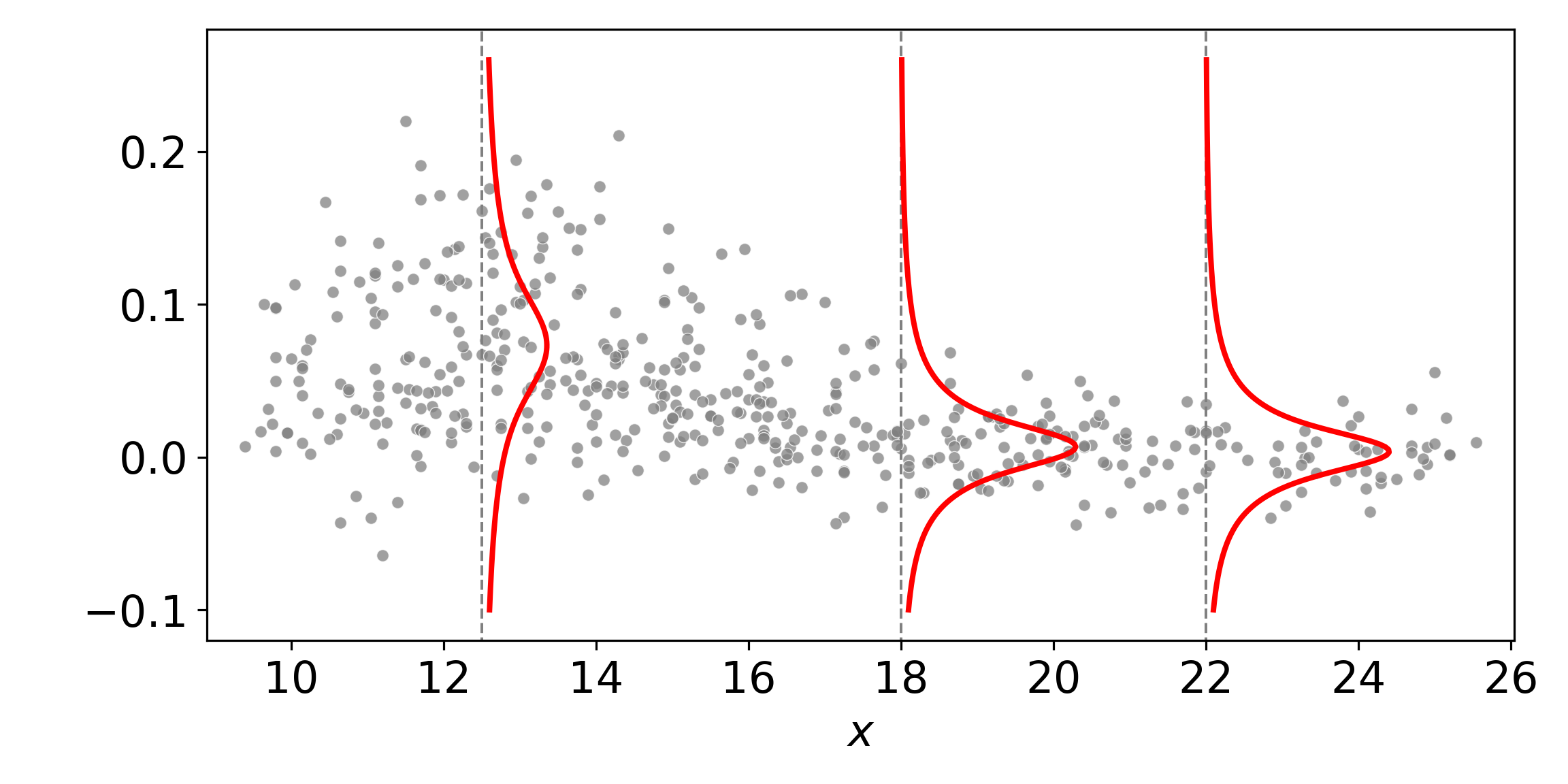}
		\hfill
		\hfill
		\caption{Conditional density estimation for the bone mineral density dataset (grey dots) by WEvidential, where the normal response distribution $\mathcal{N}(y \mid m, \sigma)$ is specified for the response variable $y$. Left: distributional estimate (10 particles) of the location parameter $\{ m^n(x)\}_{n=1}^{10}$ for each input. Right: estimated density \eqref{eq:BMA} based on the normal response distribution averaged over the output particles $\{ ( m^n(x), \sigma^n(x) ) \}_{n=1}^{10}$.}\label{fig:cde_bone}
	\end{figure}

\begin{figure}[h]
		\centering
		\hfill
		\includegraphics[width=0.425\textwidth]{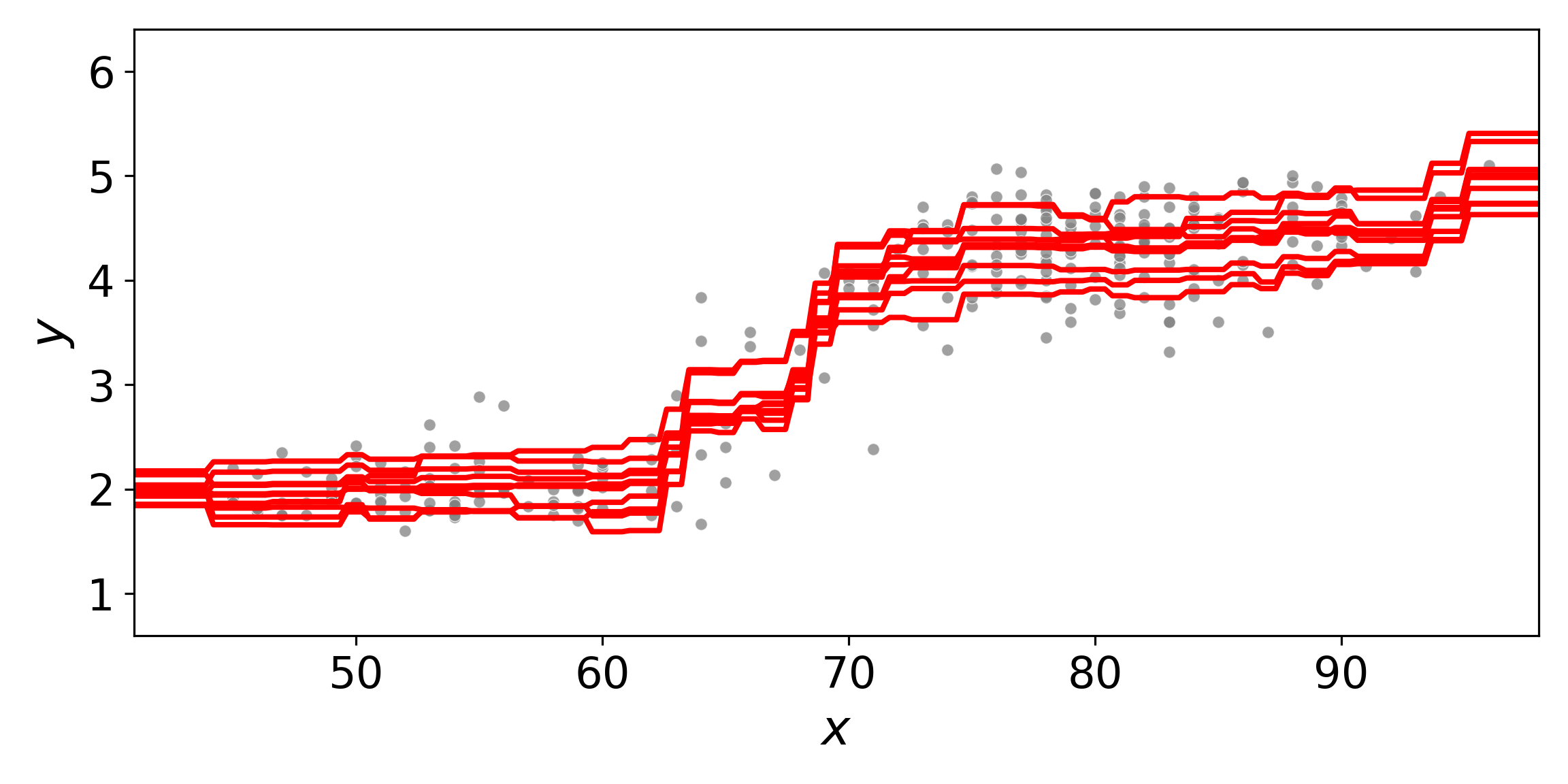}
		\hfill
		\includegraphics[width=0.425\textwidth]{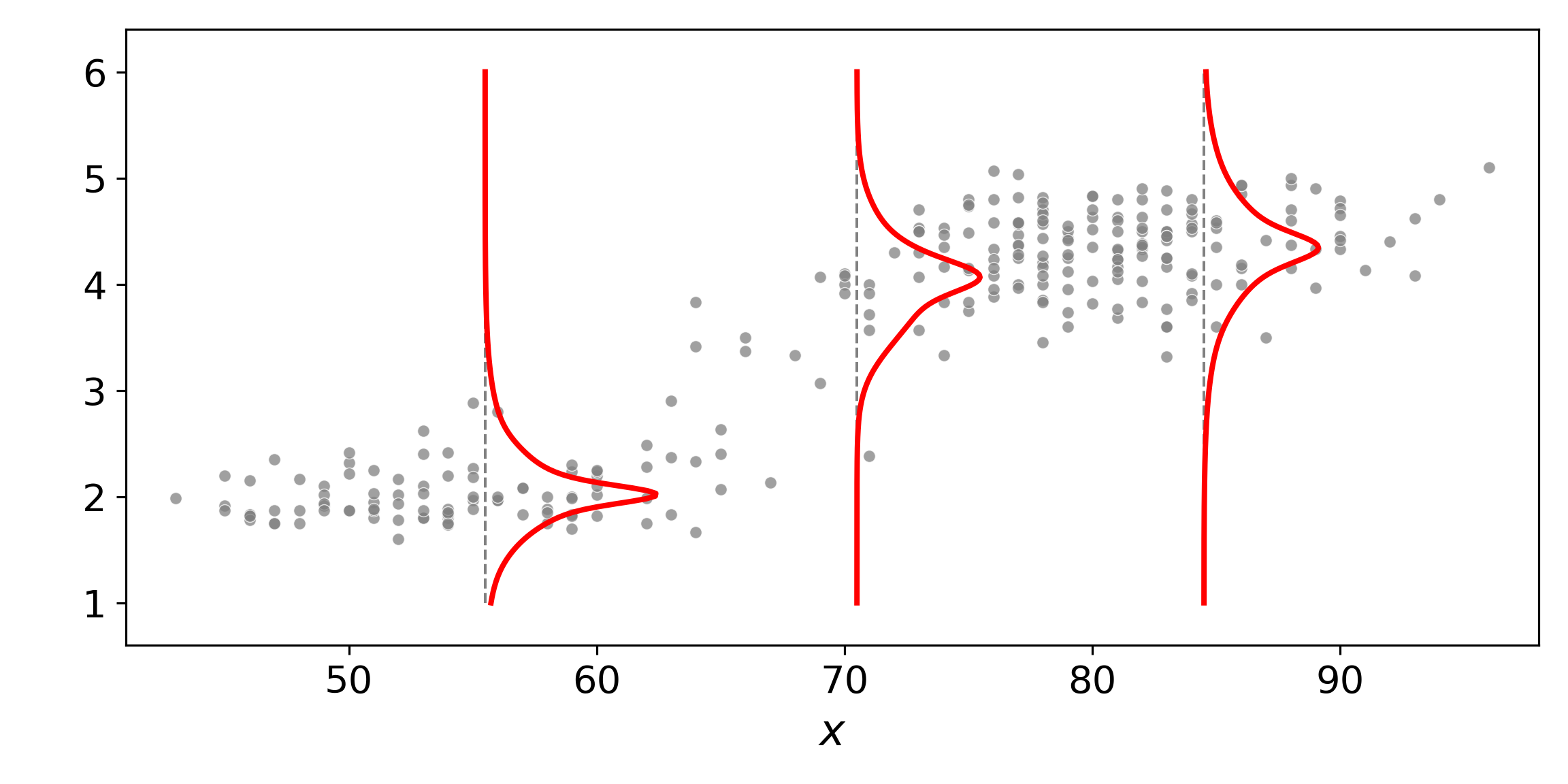}
		\hfill
		\hfill
		\caption{Conditional density estimation for the old faithful geyser dataset (grey dots) by WEvidential. Left: distributional estimate (10 particles) of the location parameter for each input. Right: estimated density by the predictive distribution \eqref{eq:BMA} based on the output particles.}\label{fig:cde_geyser}
	\end{figure}

	The conditional density is estimated using the predictive distribution \eqref{eq:BMA} by the WGBoost algorithm.
	We used two real-world datasets, \emph{bone mineral density} \cite{Hastie2009} and \emph{old faithful geyser} \cite{Weisberg1985}.
	\Cref{fig:cde_bone} depicts the result for the former dataset, demonstrating that the WGBoost algorithm captures the heterogeneity of the conditional density on each input well.
	Similarly, \Cref{fig:cde_geyser} depicts the result for the latter dataset.
	
	\subsection{Probabilistic Regression Benchmark} \label{sec:section42}
	
	This section examines the regression performance of the WGBoost algorithm using a standard benchmark protocol that originated in \cite{Hernandez-Lobato2015} and has been used in a number of subsequent works \cite{Lakshminarayanan2017,Gal2016,Duan2020}.
	The benchmark protocol uses real-world tabular datasets from the UCI machine learning repository \cite{Dua2017}, each with one-dimensional scalar responses.
	As in \Cref{sec:section41}, the normal response distribution $\mathcal{N}(y \mid m, \sigma)$ and the prior $p_i(m, \sigma)$ in \Cref{ex:normal} were used to define the individual-level posterior $p(m, \sigma \mid y_i)$.
	
	We randomly held out 10\% of each dataset as a test set, following the data splitting protocol in \cite{Hernandez-Lobato2015}.
	The negative log likelihood (NLL) is measured by using the predictive distribution \eqref{eq:BMA} by the WGBoost algorithm.
	The room mean squared error (RMSE) is measured by using the point prediction produced by taking the mean value of the predictive distribution.
	For this benchmark, we followed an approach used in \cite{Duan2020} to choose the number of weak learners $M$ by early-stopping, where we held out 20\% of the training set as a validation set to choose the number $1 \le M \le 4000$ achieving the least validation error.
	Once the number $M$ was chosen, the WGBoost algorithm was trained again using all the entire training set.
	We repeated this procedure 20 times for each dataset, except the \emph{protein} dataset for which we repeated five times.
	
	We compared the performance of WEvidential with five other methods: Monte Carlo Dropout (MCDropout) \cite{Gal2016}, Deep Ensemble (DEnsemble) \cite{Lakshminarayanan2017}, Concrete Dropout (CDropout) \cite{Gal2017}, Natural Gradient Boosting (NGBoost) \cite{Duan2020}, and Deep Evidential Regression (DEvidential) \cite{Amini2020}.
	\Cref{apx:appendix_f} briefly describes each algorithm and provides further details on the experiment.
	\Cref{tab:nll} summarises the NLLs and RMSEs of the six algorithms.
	The WGBoost algorithm achieves the best score or a score sufficiently close to the best score for the majority of the datasets.
	
	\begin{table}[t]
	\caption{The NLLs and RMSEs with the standard deviation. The best score is underlined for each dataset, and the scores whose standard deviation ranges include the best score are in bold. Results of MCDropout, DEnsembles, CDropout, NGBoost, and DEvidential were reported in \cite{Gal2016}, \cite{Lakshminarayanan2017}, \cite{Gal2017}, \cite{Duan2020} and \cite{Amini2020} respectively.}
	\label{tab:nll}
	\resizebox{\textwidth}{!}{
	\begin{tabular}{cccccccc}
		\toprule
		Dataset & Criteria & WEvidential & MCDropout & DEnsemble & CDropout & NGBoost & DEvidential  \\
		\midrule
		boston & \multirow{9}{*}{NLL} & \textbf{2.47 ± 0.16} & 2.46 ± 0.06 & \textbf{2.41 ± 0.25} & 2.72 ± 0.01 & \textbf{2.43 ± 0.15} & \underline{\textbf{2.35 ± 0.06}} \\ 
		concrete & & \underline{\textbf{2.83 ± 0.11}} & 3.04 ± 0.02 & 3.06 ± 0.18 & 3.51 ± 0.00 & 3.04 ± 0.17 & 3.01 ± 0.02 \\ 
		energy & & \underline{\textbf{0.53 ± 0.08}} & 1.99 ± 0.02 & 1.38 ± 0.22 & 2.30 ± 0.00 & \textbf{0.60 ± 0.45} & 1.39 ± 0.06 \\ 
		kin8nm & & -0.44 ± 0.03 & -0.95 ± 0.01 & -1.20 ± 0.02 & -0.65 ± 0.00 & -0.49 ± 0.02 & \underline{\textbf{-1.24 ± 0.01}} \\ 
		naval & & -5.47 ± 0.03 & -3.80 ± 0.01 & -5.63 ± 0.05 & \underline{\textbf{-5.87 ± 0.05}} & -5.34 ± 0.04 & -5.73 ± 0.07 \\ 
		power & & \underline{\textbf{2.60 ± 0.04}} & 2.80 ± 0.01 & 2.79 ± 0.04 & 2.75 ± 0.01 & 2.79 ± 0.11 & 2.81 ± 0.07 \\ 
		protein & & 2.70 ± 0.01 & 2.89 ± 0.00 & 2.83 ± 0.02 & 2.81 ± 0.00 & 2.81 ± 0.03 & \underline{\textbf{2.63 ± 0.00}} \\ 
		wine & & \textbf{0.95 ± 0.08} & 0.93 ± 0.01 & \textbf{0.94 ± 0.12} & 1.70 ± 0.00 & \textbf{0.91 ± 0.06} & \underline{\textbf{0.89 ± 0.05}} \\ 
		yacht & & \underline{\textbf{0.16 ± 0.24}} & 1.55 ± 0.03 & 1.18 ± 0.21 & 1.75 ± 0.00 & \textbf{0.20 ± 0.26} & 1.03 ± 0.19 \\ 
		\midrule
		boston & \multirow{9}{*}{RMSE} & \textbf{2.78 ± 0.60} & 2.97 ± 0.19 & \textbf{3.28 ± 1.00} & \underline{\textbf{2.65 ± 0.17}} & \textbf{2.94 ± 0.53} & 3.06 ± 0.16 \\ 
		concrete & & \underline{\textbf{4.15 ± 0.52}} & 5.23 ± 0.12 & 6.03 ± 0.58 & 4.46 ± 0.16 & 5.06 ± 0.61 & 5.85 ± 0.15 \\ 
		energy & & \underline{\textbf{0.42 ± 0.07}} & 1.66 ± 0.04 & 2.09 ± 0.29 & 0.46 ± 0.02 & \textbf{0.46 ± 0.06} & 2.06 ± 0.10 \\ 
		kin8nm & & 0.15 ± 0.00 & 0.10 ± 0.00 & 0.09 ± 0.00 & \underline{\textbf{0.07 ± 0.00}} & 0.16 ± 0.00 & 0.09 ± 0.00 \\ 
		naval & & \underline{\textbf{0.00 ± 0.00}} & 0.01 ± 0.00 & \underline{\textbf{0.00 ± 0.00}} & \underline{\textbf{0.00 ± 0.00}} & \underline{\textbf{0.00 ± 0.00}} & \underline{\textbf{0.00 ± 0.00}} \\ 
		power & & \underline{\textbf{3.19 ± 0.25}} & 4.02 ± 0.04 & 4.11 ± 0.17 & 3.70 ± 0.04 & 3.79 ± 0.18 & 4.23 ± 0.09 \\ 
		protein & & 4.09 ± 0.02 & 4.36 ± 0.01 & 4.71 ± 0.06 & \underline{\textbf{3.85 ± 0.02}} & 4.33 ± 0.03 & 4.64 ± 0.03 \\ 
		wine & & \underline{\textbf{0.61 ± 0.05}} & \textbf{0.62 ± 0.01} & \textbf{0.64 ± 0.04} & 0.62 ± 0.00 & \textbf{0.63 ± 0.04} & \underline{\textbf{0.61 ± 0.02}} \\ 
		yacht & & \underline{\textbf{0.48 ± 0.18}} & 1.11 ± 0.09 & 1.58 ± 0.48 & 0.57 ± 0.05 & \textbf{0.50 ± 0.20} & 1.57 ± 0.56 \\ 
		\bottomrule
	\end{tabular}
	}
	\end{table}

	\subsection{Classification and Out-of-Distribution Detection} \label{sec:section43}
	
	This section examines the classification and anomaly OOD detection performance of the WGBoost algorithm on two real-world tabular datasets, \emph{segment} and \emph{sensorless}, following the protocol used in \cite{Charpentier2020}.
	The categorical output distribution $\mathcal{C}(y \mid q)$ and the prior $p_i(q)$ in \Cref{ex:categorical} were used to define the individual-level posterior $p(q \mid y_i)$, in which case the output of the WGBoost algorithm is a set of $10$ particles $\{ q^n \}_{n=1}^{10}$ of the class probability parameter $q$ in the simplex $\Delta^k$ for each input $x$.
	We set the number of weak learners $M$ to $4000$.
	The dispersion of the output particles of the WGBoost algorithm was used for OOD detection \cite{Yang2024}.
	If a test input was an \emph{in-distribution} sample from the same distribution as the data, we expected the output particles to concentrate on some small region in $\Delta^k$ indicating a high probability of the correct class.
	If a test input was an OOD sample, we expected the output particles to disperse over $\Delta^k$ because the model ought to be less certain about the correct class.
	
	The segment and sensorless datasets have 7 and 11 classes in total.
	For the segment dataset, the data subset that belongs to the last class was kept as the OOD samples.
	For the sensorless dataset, the data subset that belongs to the last two classes was kept as the OOD samples.
	For each dataset, 20\% of the non-OOD samples is held out as a test set to measure the classification accuracy.
	Several approaches can define the OOD score of each input \cite{Yang2024}.
	We focused on an approach that uses the variance of the output particles as the OOD score. 
	For the WGBoost algorithm, we employed the inverse of the maximum norm of the variance as the OOD score.
	Given the OOD score, we measured the OOD detection performance by the area under the precision recall curve (PR-AUC), viewing non-OOD test data as the positive class and OOD data as the negative class.
	We repeated this procedure five times.
	
	\begin{table}[b]
		\caption{The classification accuracies and OOD detection PR-AUCs with the standard deviation. For each dataset, the best score is underlined and in bold. The results other than WEvidential were reported in \cite{Charpentier2020}.} \label{tab:ood}
		\resizebox{\textwidth}{!}{
			\begin{tabular}{ccccccc}
				\toprule
				Dataset & Criteria & WEvidential & MCDropout & DEnsemble & DDistillation & PNetwork  \\
				\midrule
				\multirow{2}{*}{segment} & Accuracy & 96.57 ± 0.6 & 95.25 ± 0.1 & \underline{\textbf{97.27 ± 0.1}} & 96.21 ± 0.1 & 96.92 ± 0.1 \\
				& OOD & \underline{\textbf{99.67 ± 0.2}} &  43.11 ± 0.6 & 58.13 ± 1.7 & 35.83 ± 0.4 &  96.74 ± 0.9 \\
				\multirow{2}{*}{sensorless} & Accuracy & \underline{\textbf{99.54 ± 0.1}} & 89.32 ± 0.2 & 99.37 ± 0.0 & 93.66 ± 1.5 & 99.52 ± 0.0 \\
				& OOD & 81.13 ± 5.3 & 40.61 ± 0.7 & 50.62 ± 0.1 &  31.17 ± 0.2 & \underline{\textbf{88.65 ± 0.4}} \\
				\bottomrule
			\end{tabular}
		}
	\end{table}
	
	We compared the WGBoost algorithm with four other methods: MCDropout, DEnsemble, and Distributional Distillation (DDistillation) \cite{Malinin2020}, and Posterior Network (PNetwork) \cite{Charpentier2020}.
	\Cref{apx:appendix_f} briefly describes each algorithm and provides further details on the experiment.
	\Cref{tab:ood} summarises the classification and OOD detection performance of the five algorithms.
	The WGBoost algorithm demonstrates a high classification and OOD detection accuracy simultaneously.
	Although PNetwork has the best OOD detection performance for the sensorless dataset, the performance of the WGBoost algorithm also exceeds 80\%, which is distinct from MCDropout, DEnsemble, and DDistillation.

	\section{Discussion}
	
	This work established the general framework of WGBoost for 'distribution-valued' supervised learning, which receives a particle-based approximation of an output distribution assigned at each input.
	We focused on application of WGBoost to evidential learning, for which we provided the setting to implement a second-order WGBoost algorithm, aligning with the standard practice in modern gradient-boosting libraries.
	We empirically demonstrated that the probabilistic forecast by WGBoost leads to better predictive accuracy and OOD detection performance.
	
	The established framework of WGBoost offers exciting avenues for future research.
	Important directions for future study include (i) investigating the convergence properties, (ii) evaluating the robustness to misspecified response distributions, and (iii) exploring alternative loss functionals to the KL divergence.
	A limitation of WGBoost may arise when data are not tabular, as in the case of standard gradient boosting.
	These questions require careful examination and are critical for future work.

	\bibliographystyle{unsrt}
	\bibliography{bibliography}

	\appendix
	\include{supplement}

\end{document}

%% file: supplement.tex
%auto-ignore
%!TEX root = main.tex

\begin{center}
	\LARGE \textbf{Appendix}
\end{center}

\vspace{40pt}

This appendix contains the technical and experiment details referred to in the main text.
\Cref{apx:appendix_a} recaps the derivation of the Wasserstein gradient and presents several examples.
\Cref{apx:appendix_b} discusses a variant of WGBoost for the KL divergence built on the unadjusted Langevin algorithm.
\Cref{apx:appendix_c} derives the diagonal approximate Wasserstein Newton direction used for WEvidential.
\Cref{apx:appendix_e} provides a simulation study to compare four different WGBoost algorithms.
\Cref{apx:appendix_f} describes the additional details of the experiment in the main text.

\section{Derivation and Example of Wasserstein Gradient} \label{apx:appendix_a}

This section recaps the derivation of the Wasserstein gradient of a functional $\F$, with examples of common divergences.
The Wasserstein gradient depends on a function on $\Theta$ called the \emph{first variation} \cite{Ambrosio2005}.
The first variation $\delta \F(\mu) / \delta \mu$ of the functional $\F$ at $\mu$ is a function on $\Theta$ that satisfies
\begin{align}
	\lim_{\epsilon \to 0^+} \frac{\F( \mu + \epsilon \nu ) - \F(\mu) }{\epsilon} = \int_\Theta \frac{\delta \F(\mu)}{\delta \mu} (\rv) \nu(\rv) d \rv
\end{align}
for all signed measure $\nu$ s.t.~$\mu + \epsilon \nu \in \P2$ for all $\epsilon$ sufficiently small.
The Wasserstein gradient $\nabla_W \F(\mu)$ of the functional $\F$ at $\mu$ is derived as the gradient of the first variation (see \cite[e.g.][]{Ambrosio2005}):
\begin{align}
	[ \nabla_W \F(\mu) ](\rv) := \nabla \frac{\delta \F(\mu)}{\delta \mu}(\rv) .
\end{align}
It is common to suppose that the functional $\F$ consists of three energies, which are determined by functions $U: \R \to \R$, $V:\Theta \to \R$, and $W:\Theta \to \R$ respectively, such that
\begin{align}
	\F(\mu) = \underbrace{ \int_\Theta U( \mu(\rv) ) d \rv }_{ \text{internal energy} } + \underbrace{ \int_\Theta V(\rv) \mu(\rv) d \rv }_{ \text{potential energy} } + \underbrace{ \frac{1}{2} \int_{\Theta \times \Theta} W(\rv - \rv') \mu(\rv) d \rv \mu(\rv') d \rv' }_{ \text{interaction energy} } . \label{eq:free_energy}
\end{align}
For a functional $\F$ that falls into the above form, the Wasserstein gradient is derived as
\begin{align}
	\left[ \nabla_W \F(\mu) \right](\rv) = \nabla U'(\mu(\rv)) + \nabla V(\rv) + \int_\Theta \nabla W(\rv - \rv') \mu(\rv') d \rv'
\end{align}
where $U'$ is the derivative of $U: \R \to \R$ \cite{Villani2003}.
The KL divergence $\F(\mu) = \KL(\mu \mid \pi)$ of a distribution $\pi$ falls into the form with $U(x) = x \log x$, $V(\rv) = - \log \pi(\rv)$, and $W(\rv) = 0$, where
\begin{align}
	\KL(\mu \mid \pi) = \int_\Theta \log \mu(\rv) \mu(\rv) d \rv + \int_\Theta - \log \pi(\rv) \mu(\rv) d \rv .
\end{align}
\Cref{table:ex_wg_div} presents examples of Wasserstein gradients of common divergences $\F(\mu) = \D(\mu \mid \pi)$.

\begin{table}[t]
	\caption{Wasserstein gradients of four divergences: the KL divergence \cite{Santambrogio2015}, the chi-squared divergence \cite{Chewi2020}, the alpha divergence \cite{Yi2023}, and the maximum mean discrepancy \cite{Arbel2019}.} \label{table:ex_wg_div}
	\centering
	\begin{tabular}{cc} 
		\toprule
		Divergence $\F(\mu) = \D(\mu \mid \pi)$ & Wasserstein gradient $\left[ \nabla_W \F(\mu) \right](\rv)$ \\
		\midrule
		$\KL(\mu \mid \pi)$ & $- ( \nabla \log \pi(\rv) - \nabla \log \mu(\rv) )$ \\ 
		$\operatorname{Chi}^2(\mu \mid \pi)$ & $2 \nabla ( \mu(\rv) / \pi(\rv) )$ \\ 
		$\operatorname{Alpha}(\mu \mid \pi)$ & $( \mu(\rv) / \pi(\rv) )^{\alpha-1} \nabla ( \mu(\rv) / \pi(\rv) )$ \\ 
		$\operatorname{MMD}(\mu \mid \pi)$ & $\int_\Theta \nabla k(\rv, \rv') \mu(\rv) d \rv - \int_\Theta \nabla k(\rv, \rv') \pi(\rv) d \rv$ \\ 
		\bottomrule
	\end{tabular}
\end{table}

In the context of Bayesian inference, the KL divergence is particularly useful among many divergences. 
The Wasserstein gradient of the KL divergence requires no normalising constant of a posterior distribution $\pi$.
This is because the Wasserstien gradient depends only on the log-gradient of the posterior $\nabla \log \pi(\rv) = \nabla \pi(\rv) / \pi(\rv)$ of the target $\pi$, in which case the normalising constant of the target $\pi$ is cancelled out by fraction.
Hence, any posterior known only up to the normalising constant can be used as the target distribution $\pi$ in the Wasserstein gradient of the KL divergence.

\section{Langevin Gradient Boosting for KL Divergence} \label{apx:appendix_b}

If a chosen functional $\F$ on $\P2$ is the KL divergence $\F(\mu) = \KL(\mu \mid \pi)$ of a target distribution $\pi$, the continuity equation \eqref{eq:continuity_equation} admits an equivalent representation as the Fokker-Planck equation \cite{Jordan1998}:
\begin{align}
	\frac{d}{d t} \mu_t = \nabla \cdot \left( \mu_t \nabla \log \pi \right) + \Delta \mu_t \quad \text{given} \quad \mu_0 \in \P2 \label{eq:fp_equation}
\end{align}
where $\Delta$ denotes the Laplacian operator. 
Recall that the original continuity equation \eqref{eq:continuity_equation} can be reformulated as the deterministic differential equation \eqref{eq:wg_update} of a random variable $\rv_t \sim \mu_t$.
In contrast, the Fokker-Planck equation \eqref{eq:fp_equation} can be reformulated as a stochastic differential equation of a random variable $\rv_t \sim \mu_t$, known as the overdamped Langevin dynamics \cite{Pavliotis2014}:
\begin{align}
	d \rv_t = \nabla \log \pi(\rv_t) d t + \sqrt{2} d B_t \quad \text{given} \quad \rv_0 \sim \mu_0 , \label{eq:ld_update}
\end{align}
where $B_t$ denotes a standard Brownian motion.
Note that the deterministic system \eqref{eq:wg_update} in the case of the KL divergence and the above stochastic system \eqref{eq:ld_update} are equivalent at population level, in a sense that the law of the random variable $\rv_t$ in both the systems solves the two equivalent equations.

At the algorithmic level, however, discretisation of each system leads to different particle update schemes.
Set the initial distribution $\mu_0$ in \eqref{eq:ld_update} to the empirical distribution $\hat{\mu}_0$ of $N$ initial particles $\{ \rv_0^n \}_{n=1}^{N}$.
Discretising the stochastic system \eqref{eq:ld_update} by the Euler-Maruyama method with a step size $\nu > 0$ yields a stochastic update scheme of particles $\{ \rv_m^n \}_{n=1}^{N}$ from step $m = 0$:
\begin{align}
	\begin{bmatrix}
		\rv_{m+1}^1 \\
		\vdots \\
		\rv_{m+1}^N
	\end{bmatrix} 
	= 
	\begin{bmatrix}
		\rv_{m}^1 \\
		\vdots \\
		\rv_{m}^N
	\end{bmatrix}
	+ \nu
	\begin{bmatrix}
		\nabla \log \pi( \rv_{m}^1 ) + \sqrt{2 / \nu}~\xi^1 \\
		\vdots \\
		\nabla \log \pi( \rv_{m}^N ) + \sqrt{2 / \nu}~\xi^N
	\end{bmatrix} , \label{eq:ld_particle_update}
\end{align}
where each $\xi^n$ denotes a realisation from a standard normal distribution on $\R^d$.
The above updating scheme of each $n$-th particle is known as the unadjusted Langevin algorithm \cite{Roberts1996}.
We can define a variant of WGBoost by replacing the term $\G_i(\mu)$ in Algorithm \ref{alg:wgb} with $ \nabla \log \mu_i( \cdot ) + \sqrt{2 / \nu}~\xi_i$ where $\mu_i$ is an output distribution at each $x_i$ and $\xi_i$ is a realisation from a standard normal distribution.
The procedure is summarised in Algorithm \ref{alg:lwgb}, which we call Langevin gradient boosting (LGBoost).

\begin{algorithm}[t]
	\caption{Langevin Gradient Boosting} \label{alg:lwgb}
	\KwIn{training set $\{ x_i, \mu_i \}_{i=1}^{D}$ of input $x_i \in \X$ and output distribution $\mu_i \in \P2$}
	\Parameter{particle number $N$, iteration $M$, rate $\nu$, weak learner $f$, initial constants $( \vartheta_0^1, \dots, \vartheta_0^N )$}
	\KwOut{set of $N$ boosting ensembles $( \B_M^1, \dots, \B_M^N )$ at final step $M$}
	$( \B_0^1(\cdot), \dots, \B_0^N(\cdot) ) \gets ( \vartheta_0^1, \dots, \vartheta_0^N )$\\
	\For{$m \gets 0, \dots, M-1$}{
		\For{$n \gets 1, \dots, N$}{
			\For{$i \gets 1, \dots, D$}{
				$g_i^n \gets \nabla \log \mu_i( \B_m^n(x_i) ) + \sqrt{2 / \nu}~\xi_i^n \quad \text{where} \quad \xi_i^n \sim \mathcal{N}(0, I_d)$
			}
			$f_{m+1}^n \gets \text{fit}\left( \left\{ x_i, g_i^n \right\}_{i=1}^{D} \right)$\\
			$\B_{m+1}^n(\cdot) \gets \B_m^n(\cdot) + \nu f_{m+1}^n(\cdot)$
		}
	}
\end{algorithm}

\section{Derivation of Approximate Wasserstein Newton Direction} \label{apx:appendix_c}

This section derives the diagonal approximate Wasserstein Newton direction based on the kernel smoothing.
The approximate Wasserstein Newton direction of the KL divergence was derived in \cite{Detommaso2018} under a different terminology---simply, the Newton direction---from a viewpoint of nonparametric variational inference.
We place their result in the context of approximate Wasserstein gradient flows.
\Cref{apx:appendix_c1} shows the derivation of the smoothed Wasserstein gradient and Hessian.
\Cref{apx:appendix_c2} defines the Newton direction built upon the smoothed Wasserstein gradient and Hessian, following the derivation in \cite{Detommaso2018}.
\Cref{apx:appendix_c3} derives the diagonal approximation of the Newton direction.

\subsection{Smoothed Wasserstein Gradient and Hessian} \label{apx:appendix_c1}

Consider the one-dimensional case $\Theta = \R$ for simplicity.
For a map $T: \R \to \R$ and a distribution $\mu \in \P2$, let $\mu_t$ be the pushforward of $\mu$ under the transform $\rv \mapsto \rv + t T(\rv)$ defined with a time-variable $t \in \R$.
This means that $\mu_t$ is a distribution obtained by change-of-variable applied for $\mu$.
The Wasserstein gradient of a functional $\F(\mu)$ can be associated with the time derivative $(d / d t) \F(\mu_t)$ \cite{Villani2003}.
In what follows, we focus on the KL divergence $\F(\mu) = \KL(\mu \mid \pi)$ as a loss functional.
Under a condition $T \in L^2(\mu)$, the time derivative at $t = 0$ satisfies the following equality
\begin{align}
	\frac{d}{d t} \KL(\mu_t \mid \pi) \Big|_{t=0} = \int_\Theta ~ T(\rv) \left[ \G^{\KL}(\mu) \right](\rv) ~ d \mu(\rv) = \left\langle T, \G^{\KL}(\mu) \right\rangle_{L^2(\mu)} , \label{eq:wg_timederiv}
\end{align}
where $\G^{\KL}(\mu)$ denotes the Wasserstein gradient of $\F(\mu) = \KL(\mu \mid \pi)$ with the target distribution $\pi$ made implicit.
It gives an interpretation of the Wasserstein gradient as the steepest-descent direction because the decay of the KL divergence at $t = 0$ is maximised when $T = - \G^{\KL}(\mu)$.

The `smoothed' Wasserstein gradient can be derived by restricting the transform map $T$ to a more regulated Hilbert space than $L^2(\mu)$.
A reproducing kernel Hilbert space (RKHS) $H$ associated with a kernel function $k: \R \times \R \to \R$ is the most common choice of such a Hilbert space \cite[e.g.][]{Liu2017}.
An important property of the RKHS $H$ is that any function $f \in H$ satisfies the \emph{reproducing property} $f(\rv) = \langle f(\cdot), k(\cdot, \rv) \rangle_{H}$ under the associated kernel $k$ and inner product $\langle \cdot, \cdot \rangle_{H}$ \cite{Paulsen2016}.
As discussed in \cite[e.g.][]{Korba2020}, applying the reproducing property in \eqref{eq:wg_timederiv} under the condition $T \in H$ and exchanging the integral order, the time derivative satisfies an alternative equality as follows:
\begin{align}
	\frac{d}{d t} \KL(\mu_t \mid \pi) \Big|_{t=0} & = \int_\Theta ~ \left\langle T(\cdot), k(\cdot, \rv) \right\rangle_{H} \left[ \G^{\KL}(\mu) \right](\rv) ~ d \mu(\rv) \\
	& = \left\langle T(\cdot), \int_\Theta \left[ \G^{\KL}(\mu) \right](\rv) k(\cdot, \rv) d \mu(\rv) \right\rangle_{H} = \left\langle T, \G^*(\mu) \right\rangle_{H} \label{eq:first_derivative_kl}
\end{align}
where $[ \G^*(\mu) ](\cdot) := \int_\Theta [ \G^{\KL}(\mu) ](\rv) k(\cdot, \rv) d \mu(\rv)$ corresponds to the smoothed Wasserstein gradient used in the main text.
The decay of the KL divergence at $t = 0$ is maximised by $T = - \G^*(\mu)$.

Similarly, the Wasserstein Hessian of the functional $\F(\mu)$ can be associated with the second time derivative $(d^2 / d t^2) \F(\mu_t)$ \cite{Villani2003}.
As discussed in \cite[e.g.][]{Korba2020}, the Wasserstein Hessian of the KL divergence, denoted $\Hess(\mu)$, is an operator over functions $T \in L^2(\mu)$ that satisfies
\begin{align}
	\frac{d^2}{d t^2} \KL(\mu_t \mid \pi) \Big|_{t=0} & = \left\langle T, \Hess(\mu) T \right\rangle_{L^2(\mu)} . \label{eq:wg_secondtimederiv}
\end{align}
See \cite{Korba2020} for the explicit form of the Wasserstein Hessian.
In the same manner as the smoothed Wasserstein gradient, applying the reproducing property in \eqref{eq:wg_secondtimederiv} under the condition $T \in H$ and exchanging the integral order, the second time derivative satisfies an alternative equality as follows:
\begin{align}
	\frac{d^2}{d t^2} \KL(\mu_t \mid \pi) \Big|_{t=0} & = \Big\langle T(\star_1), \big\langle \left[ \Hess^*(\mu) \right](\star_1, \star_2) , T(\star_2) \big\rangle_{H} \Big\rangle_{H} \label{eq:second_derivative_kl}
\end{align}
where $[ \Hess^*(\mu) ](\star_1, \star_2) := \langle k(\star_1, \cdot), \Hess(\mu) k(\star_2, \cdot) \rangle_{L^2(\mu)}$ is the smoothed Wasserstein Hessian and the symbols $\star_1$ and $\star_2$ denote the variables to which each of the two inner products is taken.

In the multidimensional case $\Theta = \R^d$, the transport map $T$ is a vector-valued function $T: \R^d \to \R^d$, where a similar derivation can be repeated by replacing $L^2(\mu)$ and $H$ with the product space of $d$ independent copies of $L^2(\mu)$ and $H$.
It follows from Proposition 1 and Theorem 1 in \cite{Detommaso2018}---which derives the explicit form of \eqref{eq:first_derivative_kl} and \eqref{eq:second_derivative_kl} under their terminology, first and second variations---that the explicit form of the smoothed Wasserstein gradient and Hessian is given by
\begin{align}
	\left[ \G^*(\mu) \right]( \cdot ) & = \E_{\rv \sim \mu}\Big[ - \nabla \log \pi(\rv) k(\cdot, \rv) - \nabla k(\cdot, \rv) \Big] \in \R^d , \label{eq:SWG} \\
	\left[ \Hess^*(\mu) \right](\star_1, \star_2) & = \E_{\rv \sim \mu}\Big[ - \nabla^2 \log \pi(\rv) k(\star_1, \rv) k(\star_2, \rv) + \nabla k(\star_1, \rv) \otimes \nabla k(\star_2, \rv) \Big] \in \R^{d \times d} \label{eq:SWH}
\end{align}
where $\nabla^2$ denotes an operator to take the Jacobian of the gradient---i.e., $\nabla^2 f(\rv)$ is the Hessian matrix of $f$ at $\rv$---and $\otimes$ denotes the outer product of two vectors.
Note that both the smoothed Wasserstein gradient and Hessian are well-defined for any distribution $\mu$ including empirical distributions.

\subsection{Approximate Wasserstein Newton Direction} \label{apx:appendix_c2}

In the Euclidean space, the Newton direction of an objective function is a direction s.t.~the second-order Taylor approximation of the function is minimised.
Similarly, \cite{Detommaso2018} characterised the Newton direction $T^*: \R^d \to \R^d$ of the KL divergence $\KL(\mu \mid \pi)$ as a solution of the following equation
\begin{align}
	\Big\langle \big\langle [\Hess^*(\mu)](\star_1, \star_2), T^*(\star_2) \big\rangle_{H} + [\G^*(\mu)](\star_1) , V(\star_1) \Big\rangle_{H} = 0 \quad \text{for all} \quad V \in H . \label{eq:WND_characterisation}
\end{align}
Here $\Theta = \R^d$ and $H$ is the product space of $d$ independent copies of the RKHS of a kernel $k$.
To obtain a closed-form solution, \cite{Detommaso2018} supposed that the Newton direction $T^*$ can be expressed in a form $T^*(\cdot) = \sum_{i=1}^{n} W^n k(\cdot, \rv^n)$ dependent on a set of each particle $\rv^n \in \Theta$ and associated vector-valued coefficient $W^n \in \R^d$.
Once the set of the particles is given, the set of the associated vector-valued coefficients is determined by solving the following simultaneous linear equation
\begin{align}
	\begin{bmatrix}
		\sum_{n=1}^{N} [\Hess^*(\mu)](\rv^1, \rv^n) \cdot W^n \\
		\vdots \\
		\sum_{n=1}^{N} [\Hess^*(\mu)](\rv^N, \rv^n) \cdot W^n
	\end{bmatrix}
	=
	\begin{bmatrix}
		- [\G^*(\mu)](\rv^1) \\
		\vdots \\
		- [\G^*(\mu)](\rv^N)
	\end{bmatrix} . \label{eq:optimal_coef_eq}
\end{align}
These equations \eqref{eq:optimal_coef_eq} can be rewritten in a matrix form \cite{Leviyev2022}.
Let $K := N \times d$.
Define a block matrix $\mathbf{H} \in \R^{K \times K}$ and a block vector $\mathbf{G} \in \R^{K}$ by the following partitioning
\begin{align}
	\mathbf{H} = \left( \begin{array}{c | c | c}
		\mathbf{H}_{11} & \cdots & \mathbf{H}_{1N} \\ \hline
		\vdots & \ddots & \vdots \\ \hline 
		\mathbf{H}_{N1} & \cdots & \mathbf{H}_{NN}
	\end{array} \right) 
	\qquad \text{and} \qquad
	\mathbf{G} = \left( \begin{array}{c}
		\mathbf{G}_{1} \\ \hline
		\vdots \\ \hline 
		\mathbf{G}_{N}
	\end{array} \right)
\end{align}
with each block specified as $\mathbf{H}_{ij} := [\Hess^*(\mu)](\rv^i, \rv^j) \in \R^{d \times d}$ and $\mathbf{G}_{i} := [\G^*(\mu)](\rv^i) \in \R^d$.
Define a block matrix $\mathbf{K} \in \R^{K \times K}$ and a block vector $\mathbf{W} \in \R^{K}$ by the following partitioning
\begin{align}
	\mathbf{K} := \left( \begin{array}{c | c | c}
		\mathbf{K}_{11} & \cdots & \mathbf{K}_{1N} \\ \hline
		\vdots & \ddots & \vdots \\ \hline 
		\mathbf{K}_{N1} & \cdots & \mathbf{K}_{NN}
	\end{array} \right) 
	\qquad \text{and} \qquad
	\mathbf{W} := \left( \begin{array}{c}
		W^1 \\ \hline
		\vdots \\ \hline 
		W^N
	\end{array} \right)
\end{align}
with each block of $\mathbf{K}$ specified as $\mathbf{K}_{ij} := \mathbf{I}_{d} \times k(\rv^i, \rv^j) \in \R^{d \times d}$, where $\mathbf{I}_{d}$ denotes the $d \times d$ identity matrix.
Notice that $\mathbf{W}$ is a block vector that aligns the vector-valued coefficients $\{ W^n \}_{n=1}^{N}$.
Using these notations, the optimal coefficients that solve \eqref{eq:optimal_coef_eq} is simply written as $\mathbf{W} = - \mathbf{H}^{-1} \mathbf{G}$ \cite{Leviyev2022}.

Given the optimal coefficients $\mathbf{W} = - \mathbf{H}^{-1} \mathbf{G}$, the Newton direction $T^*(\rv^n)$ evaluated at the given particle $\rv^n$ for each $n = 1, \dots, N$ can be written in the following block vector form
\begin{align}
	\left( \begin{array}{c}
		T^*(\rv^1) \\ \hline
		\vdots \\ \hline
		T^*(\rv^N)
	\end{array} \right)
	= - 
	\left( \begin{array}{c | c | c}
		\mathbf{K}_{11} & \cdots & \mathbf{K}_{1N} \\ \hline
		\vdots & \ddots & \vdots \\ \hline 
		\mathbf{K}_{N1} & \cdots & \mathbf{K}_{NN}
	\end{array} \right)
	\left( \begin{array}{c | c | c}
		\mathbf{H}_{11} & \cdots & \mathbf{H}_{1N} \\ \hline
		\vdots & \ddots & \vdots \\ \hline 
		\mathbf{H}_{N1} & \cdots & \mathbf{H}_{NN}
	\end{array} \right)^{-1}
	\left( \begin{array}{c}
		\mathbf{G}_{1} \\ \hline
		\vdots \\ \hline 
		\mathbf{G}_{N}
	\end{array} \right) \label{eq:swf_newton}
\end{align}
To distinguish from the standard Newton direction in the Euclidean space, we call \eqref{eq:swf_newton} the approximate Wasserstein Newton direction.
The approximate Wasserstein Newton direction yields a second-order particle update scheme.
Suppose we have particles $\{ \rv_m^n \}_{n=1}^{N}$ to be updated at each step $m$.
At each step $m$, define the above matrices $\mathbf{H}$ and $\mathbf{G}$ with the empirical distribution $\mu = \hat{\pi}_m$ of the particles $\{ \rv_m^n \}_{n=1}^{N}$.
Replacing the Wasserstein gradient in the particle update scheme \eqref{eq:wg_particle_update} by the approximate Wasserstein Newton direction \eqref{eq:swf_newton} provides the second-order update scheme in \cite{Detommaso2018}.

\subsection{Diagonal Approximate Wasserstein Newton Direction} \label{apx:appendix_c3}

We derive the diagonal approximation of the approximate Wasserstein Newton direction, which we used for our second-order WGBoost algorithm.
A few approximations of the approximate Wasserstein Newton direction were discussed in \cite{Detommaso2018} for better performance of their particle algorithm.
We derive the diagonal approximation so that no matrix product and inversion will be involved.
Specifically, we replace the matrices $\mathbf{K}$ and $\mathbf{H}$ in \eqref{eq:swf_newton} by the diagonal approximations $\hat{\mathbf{K}}$ and $\hat{\mathbf{H}}$, that is,
\begin{align}
	\hat{\mathbf{K}} = \left( \begin{array}{c | c | c}
		\mathbf{I}_{d} & \cdots & \mathbf{0} \\ \hline
		\vdots & \ddots & \vdots \\ \hline 
		\mathbf{0} & \cdots & \mathbf{I}_{d}
	\end{array} \right) 
	\quad \text{and} \quad
	\hat{\mathbf{H}} = \left( \begin{array}{c | c | c}
		\mathbf{h}_{11} & \cdots & \mathbf{0} \\ \hline
		\vdots & \ddots & \vdots \\ \hline 
		\mathbf{0}  & \cdots & \mathbf{h}_{NN}
	\end{array} \right) ,
\end{align}
where $\mathbf{K}_{nn} = \mathbf{I}_d \times k(\rv^n, \rv^n) = \mathbf{I}_d$ for the Gaussian kernel $k$ used in this work, and the matrix $\mathbf{h}_{nn} \in \R^{d \times d}$ denotes the diagonal approximation of the diagonal block $\mathbf{H}_{nn}$ of $\mathbf{H}$.

Recall that $\mathbf{H}_{nn} = [\Hess^*(\mu)](\rv^n, \rv^n)$. 
Denote by $\operatorname{Diag}(\mathbf{A})$ the diagonal of a square matrix $\mathbf{A}$.
The diagonal approximation $\mathbf{h}_{nn}$ is a diagonal matrix whose diagonal is $\operatorname{Diag}(\mathbf{H}_{nn})$. 
We plug the diagonal approximations $\hat{\mathbf{K}}$ and $\hat{\mathbf{H}}$ in \eqref{eq:swf_newton}.
It follows from inverse and multiplication properties of diagonal matrices that the approximate Wasserstein Newton direction turns into a form
\begin{align}
	\left( \begin{array}{c}
		T^*(\rv^1) \\ \hline
		\vdots \\ \hline
		T^*(\rv^N)
	\end{array} \right)
	= - 
	\left( \begin{array}{c | c | c}
		\mathbf{h}_{11} & \cdots & \mathbf{0} \\ \hline
		\vdots & \ddots & \vdots \\ \hline 
		\mathbf{0}  & \cdots & \mathbf{h}_{NN}
	\end{array} \right)^{-1}
	\left( \begin{array}{c}
		\mathbf{G}_{1} \\ \hline
		\vdots \\ \hline
		\mathbf{G}_{N}
	\end{array} \right)
	=
	\left( \begin{array}{c}
		- \mathbf{G}_{1} \oslash \operatorname{Diag}\left( \mathbf{H}_{11} \right) \\ \hline
		\vdots \\ \hline
		- \mathbf{G}_{N} \oslash \operatorname{Diag}\left( \mathbf{H}_{NN} \right) 
	\end{array} \right) . \label{eq:swf_newton_diagonal}
\end{align}
At an arbitrary particle location $\rv$, denote by $[ \H^*(\mu) ](\rv)$ the diagonal of the smoothed Wasserstein Hessian $[\Hess^*(\mu)](\rv, \rv)$.
It is straightforward to see that the diagonal can be written as
\begin{align}
	\left[ \H^*(\mu) \right]( \cdot ) = \E_{\rv \sim \mu}\Big[ - \nabla_{\text{d}}^2 \log \pi(\rv) k(\cdot, \rv)^2 + \nabla k(\cdot, \rv) \odot \nabla k(\cdot, \rv) \Big] . \label{eq:DSWH}
\end{align}
Notice that $\operatorname{Diag}\left( \mathbf{H}_{nn} \right) = \left[ \H^*(\mu) \right]( \rv^n )$ by definition.
It therefore follows from the formula \eqref{eq:swf_newton_diagonal} with $\mathbf{G}_{n} = [ \G^*(\mu) ]( \rv^n )$ and $\operatorname{Diag}\left( \mathbf{H}_{nn} \right) = \left[ \H^*(\mu) \right]( \rv^n )$ that the diagonal approximate Wasserstein Newton direction at an arbitrary particle location $\rv$ can be independently computed by
\begin{align}
	- [\G^*(\mu)](\rv) \oslash \left[ \H^*(\mu) \right](\rv) .
\end{align}
We used this direction in \Cref{sec:section3}.
In the main text, the diagonal approximate Wasserstein Newton direction is defined for each loss functional $\F_i(\cdot) = \D(\cdot \mid \mu_i)$, with $\pi = \mu_i$, using the smoothed Wasserstein gradient $\G_i^*(\mu)$ and the diagonal of the smoothed Wasserstein Hessian $\H_i^*(\mu)$ for each i-th output distribution $\mu_i$.

\section{Comparison of Different WGBoost Algorithms} \label{apx:appendix_e}

We compare four different algorithms of WGBoost for the KL divergence through a simulation study.
The first three algorithms are defined by setting the term $\G_i(\mu)$ in Algorithm \ref{alg:wgb} to, respectively,
\begin{enumerate}
	\item the smoothed Wasserstein gradient in \eqref{eq:swg};
	\item the diagonal approximate Wasserstein Newton direction in \eqref{eq:sswg};
	\item the full approximate Wasserstein Newton direction in \eqref{eq:swf_newton}.
\end{enumerate}
The fourth algorithm, which is rather a variant of WGBoost, is LGBoost in \Cref{apx:appendix_b}.
The first and third WGBoost algorithms is called, respectively, first-order WEvidential and full-Newton WEvidential.
The second WGBoost algorithm is WEvidential presented in \Cref{sec:section3}.
We fit the four algorithms to a synthetic dataset $\{ x_i, \mu_i \}_{i=1}^{D}$ whose inputs are 200 gird points on the interval $[- 3.5, 3.5]$ and output distributions are normal distributions $\mu_i(\rv) = \mathcal{N}( \rv \mid \text{sin}(x_i), 0.5 )$ conditional on each $x_i$.

The first-order WEvidential is implemented by Algorithm \ref{alg:wel} removing $h_i^n$ and replacing $g_i^n \oslash h_i^n$ with $g_i^n$.
The full-Newton WEvidential is implemented by Algorithm \ref{alg:wel} replacing $g_i^n \oslash h_i^n$ with $v_i^n$ computed by the following Algorithm \ref{alg:awnd}, where $\nabla^2 f(\theta)$ denotes the Hessian matrix of a function $f: \Theta \to \R$ at $\theta$.

\begin{algorithm}[h]
	\caption{Computation of Approximate Wasserstein Newton Direction} \label{alg:awnd}
	\KwIn{input $x_i$, output distribution $\mu_i$, outputs of $N$ boostings $( F_m^1(x_i), \dots, F_m^N(x_i) )$ for input $x_i$}
	\KwOut{Wasserstein Newton direction $(v_i^1, \dots, v_i^N)$ evaluated at $( F_m^1(x_i), \dots, F_m^N(x_i) )$ for input $x_i$}
	$\hat{\mu}_{m,i} \gets \text{empirical distribution of set of $N$ outputs } (\B_m^1(x_i), \dots, \B_m^N(x_i))$\text{ for input $x_i$}\\
	\For{$n \gets 1, \dots, N$}{
		$g_i^n \gets \E_{\rv^* \sim \hat{\mu}_{m,i}}[ \nabla \log \mu(\rv^*) k(\B_m^n(x_i), \rv^*) + \nabla k(\B_m^n(x_i), \rv^*) ]$\\
		\For{$k \gets 1, \dots, N$}{
			$H_i^{nk} \gets \E_{\rv^* \sim \hat{\mu}_{m,i}}[ - \nabla^2 \log \mu_i(\rv^*) k(\B_m^n(x_i), \rv^*) k(\B_m^k(x_i), \rv^*) + \nabla k(\B_m^n(x_i), \rv^*) \otimes \nabla k(\B_m^k(x_i), \rv^*) ]$\\
			$K_i^{nk} \gets \mathbf{I}_d \cdot k(\B_m^n(x_i), \B_m^k(x_i))$
		}
	}
	$\left( \begin{array}{c}
		v_i^1 \\ \hline
		\vdots \\ \hline
		v_i^N
	\end{array} \right)
	\gets
	\left( \begin{array}{c | c | c}
		K_i^{11} & \cdots & K_i^{1N} \\ \hline
		\vdots & \ddots & \vdots \\ \hline 
		K_i^{N1} & \cdots & K_i^{NN}
	\end{array} \right)
	\left( \begin{array}{c | c | c}
		H_i^{11} & \cdots & H_i^{1N} \\ \hline
		\vdots & \ddots & \vdots \\ \hline 
		H_i^{N1} & \cdots & H_i^{NN}
	\end{array} \right)^{-1}
	\left( \begin{array}{c}
		g_i^{1} \\ \hline
		\vdots \\ \hline 
		g_i^{N}
	\end{array} \right)$
\end{algorithm}

\begin{figure}[h]
	\centering
	\includegraphics[width=0.49\textwidth]{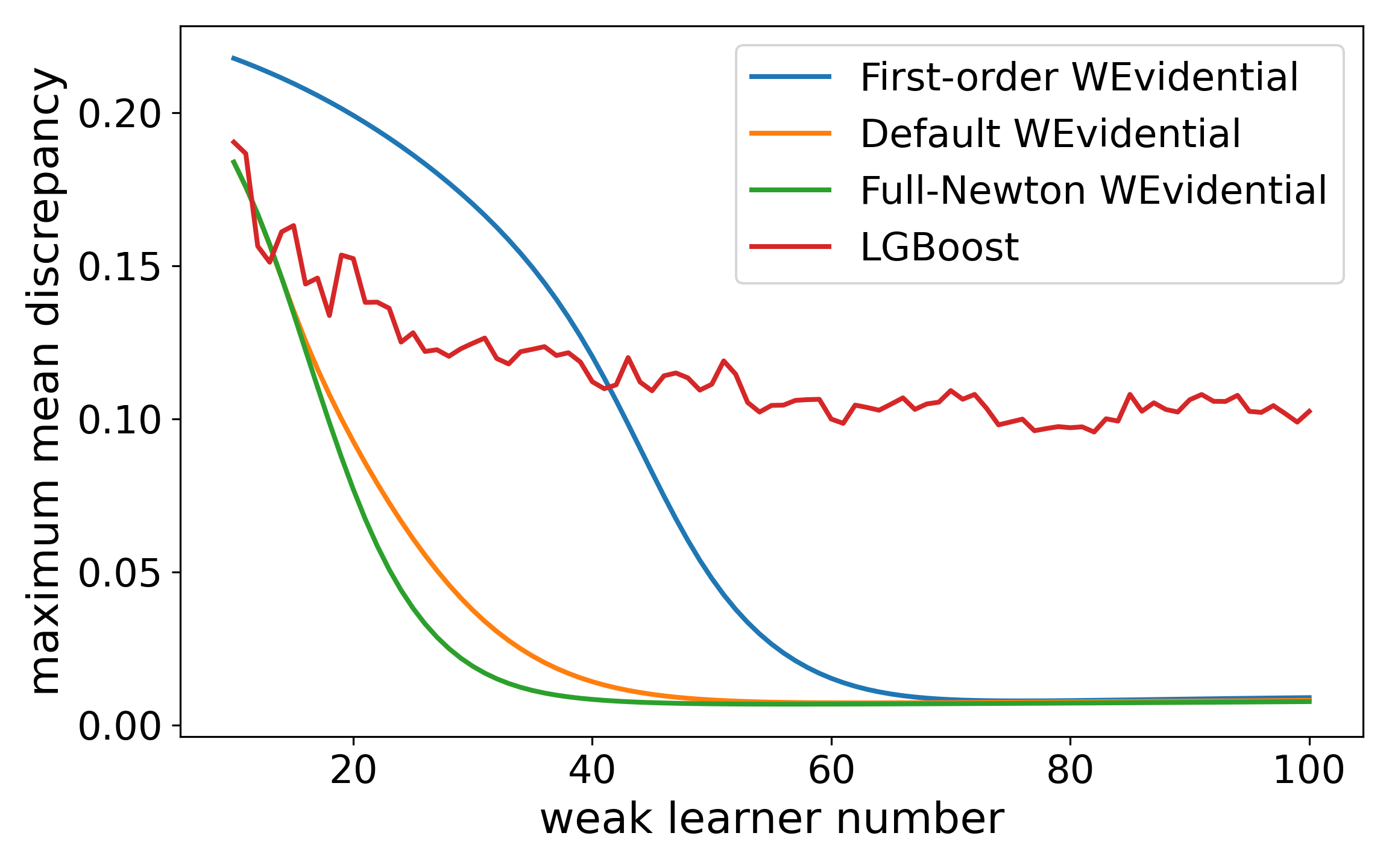}
	\hfill
	\hfill
	\includegraphics[width=0.49\textwidth]{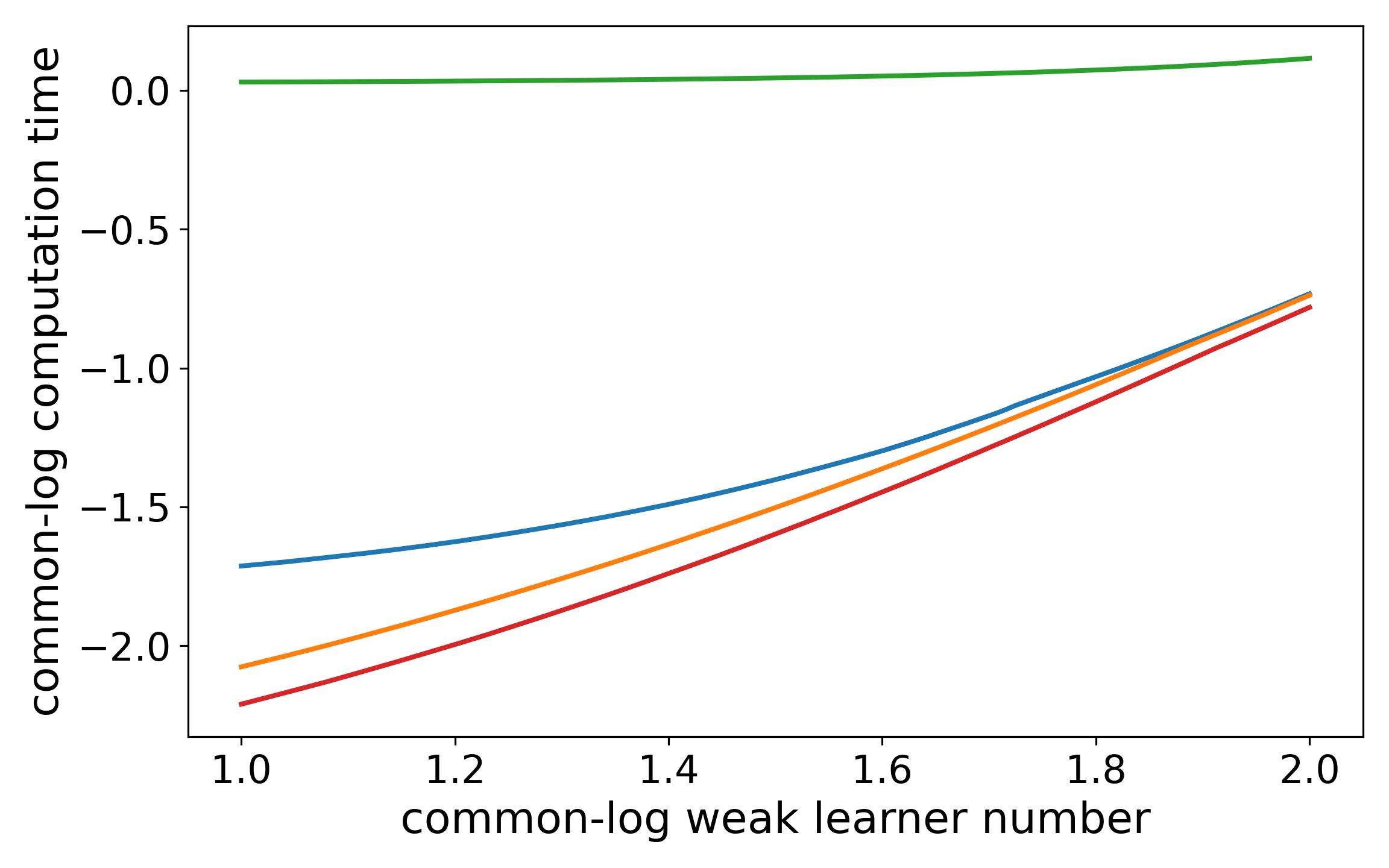}
	\caption{The approximation error and computational time of the four algorithms. Left: approximation error of each algorithm measured by the MMD averaged over the inputs. Right: computational time with respect to the weak learner number in common logarithm scale.}\label{fig:supplement_result}
\end{figure}

\Cref{fig:supplement_result} shows the performance and computational time of each algorithm on the synthetic data with respect to the number of weak learners.
We computed the output of each algorithm for 500 grid points in the interval $[-3.5, 3.5]$.
We used the maximum mean discrepancy (MMD) \cite{Smola2007} to measure the approximation error between the empirical distribution $\hat{\mu}_i$ of the output particles and the output distribution $\mu_i$ at each input $x_i$:
\begin{align}
	\operatorname{MMD}^2\left( \hat{\mu}_i, \mu_i \right) = \E_{\rv \sim \hat{\mu}_i, \rv' \sim \hat{\mu}_i}[ k(\rv, \rv') ] - 2 \E_{\rv \sim \hat{\mu}_i, \rv' \sim \mu_i}[ k(\rv, \rv') ] + \E_{\rv \sim \mu_i, \rv' \sim \mu_i}[ k(\rv, \rv') ]
\end{align}
where $k$ is a Gaussian kernel $k(\rv, \rv') = \exp( - (\rv - \rv')^2 / h)$ with scale hyperparameter $h = 0.025$.
The total approximation error was measured by the MMD averaged over all the inputs.
We set the initial constant $\{ \vartheta^n \}_{n=1}^{10}$ of each algorithm to 10 grid points in the interval $[-10, 10]$, which sufficiently differs from the output distributions to observe the decay of the approximation error.
The decision tree regressor with maximum depth $3$ was used as weak learners for all the algorithm.

\begin{figure}[h]
	\centering
	\hfill
	\subcaptionbox{First-order WEvidential}{\includegraphics[width=0.24\textwidth]{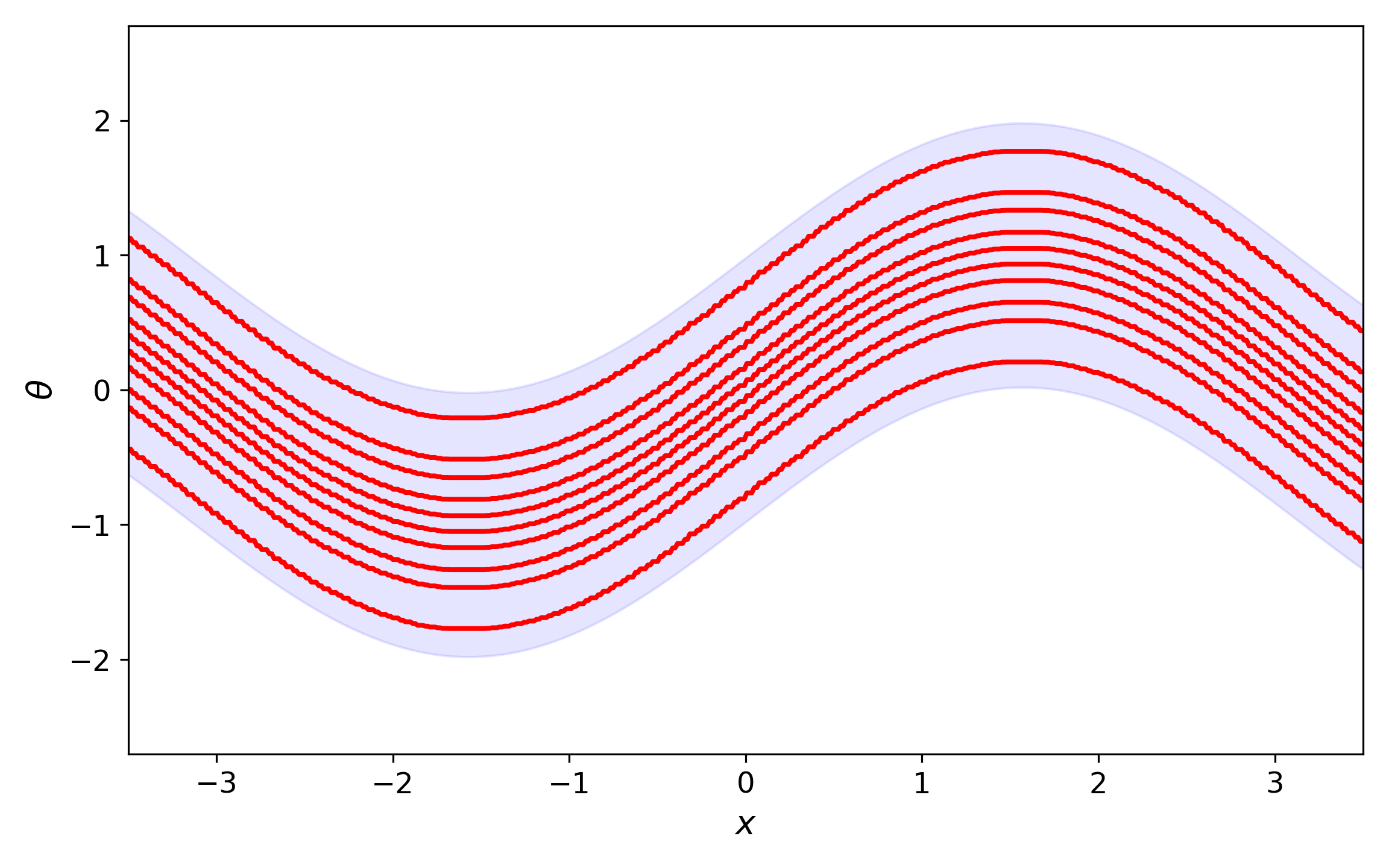}}
	\hfill
	\subcaptionbox{WEvidential}{\includegraphics[width=0.24\textwidth]{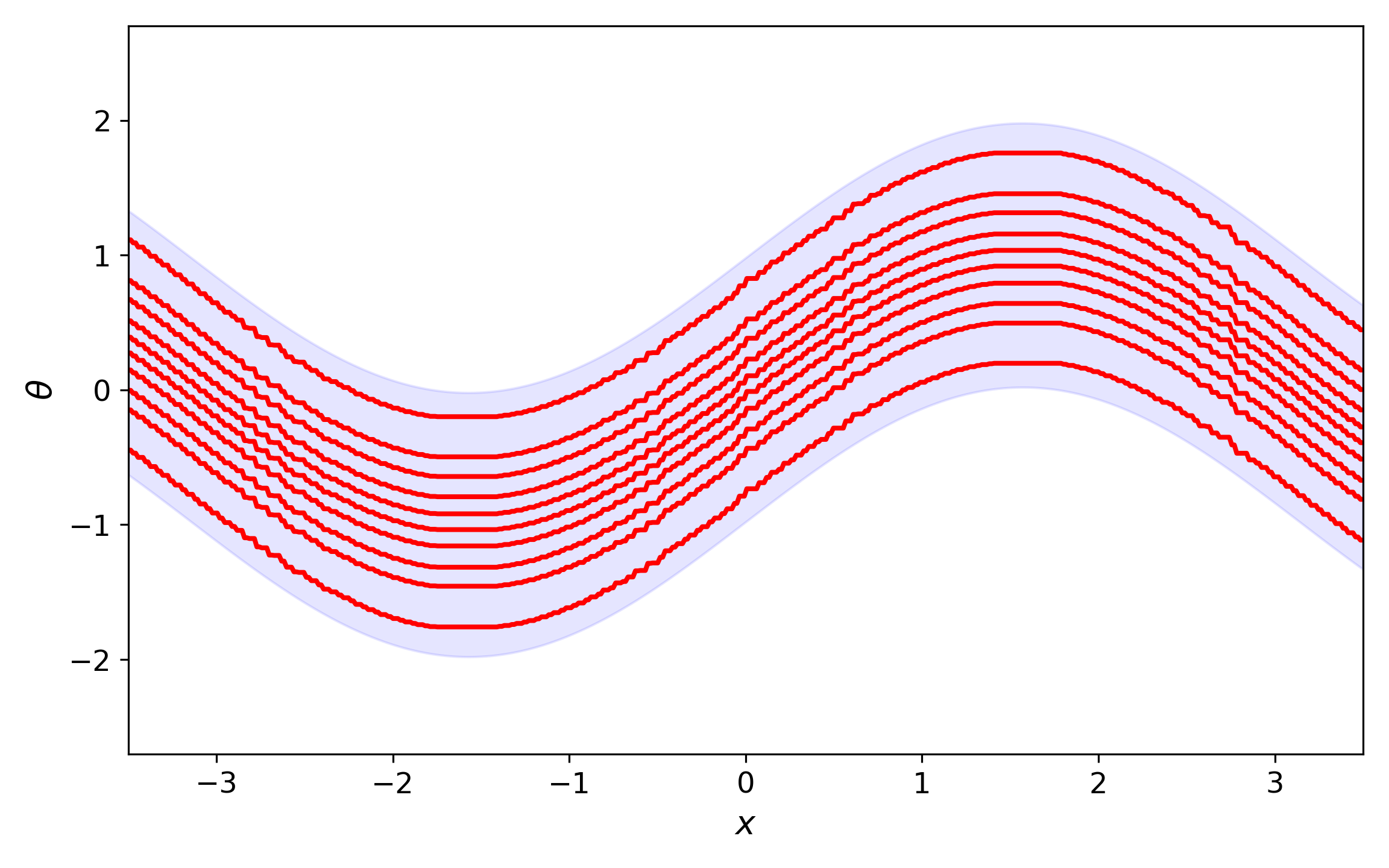}}
	\hfill
	\subcaptionbox{Full-Newton WEvidential}{\includegraphics[width=0.24\textwidth]{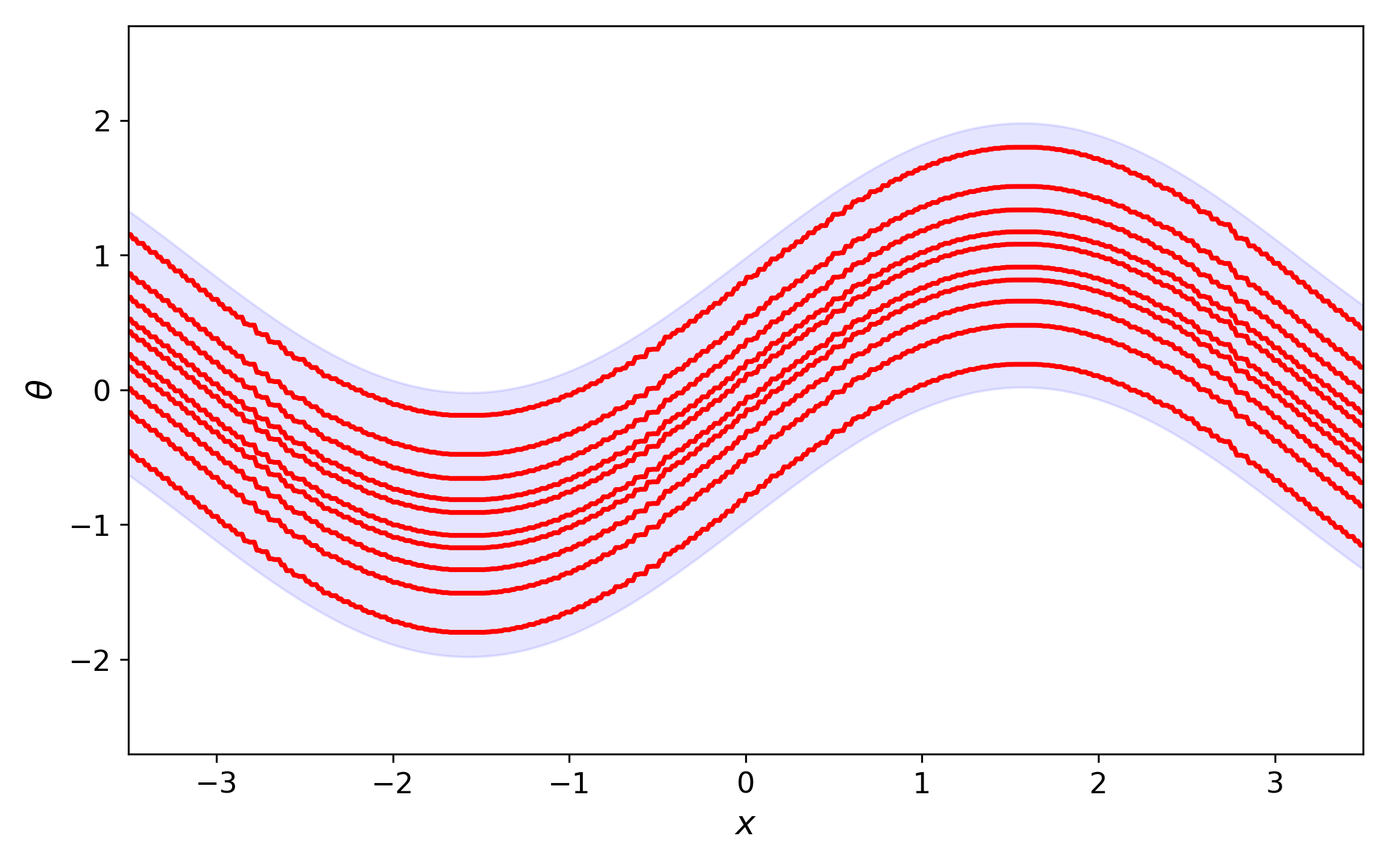}}
	\hfill
	\subcaptionbox{LGBoost}{\includegraphics[width=0.24\textwidth]{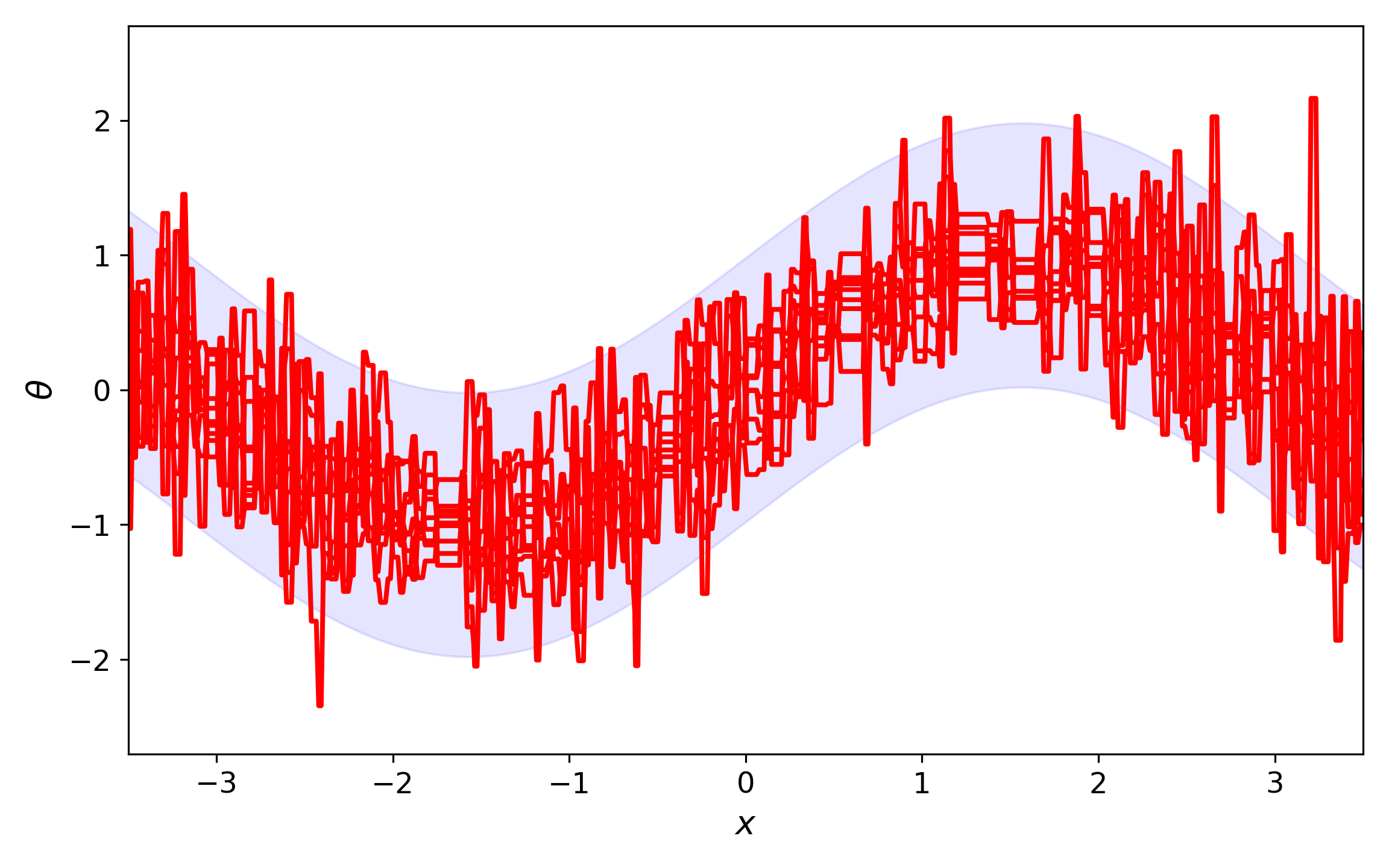}}
	\hfill
	\hfill
	\caption{Illustration of the output of the four algorithms (red line) with 100 weak learners trained. The blue area corresponds to the 95\% high probability region of each output distribution.}\label{fig:supplement_output}
\end{figure}

\Cref{fig:supplement_result} demonstrates that WEvidential and full-Newton WEvidential reduce the approximation error most efficiently, while full-Newton WEvidential takes the longest computational time among others.
As in Algorithm \ref{alg:awnd}, the computation of the full approximate Wasserstein Newton direction requires the inverse and product of the $(N \times d) \times (N \times d)$ block matrices, where $N$ denotes the particle number $N$ and $d$ denotes the particle dimension.
The computation of the diagonal approximate Wasserstein Newton direction requires only elementwise division of the $d$-dimensional vectors.
The error decay of LGBoost is not as fast as the other WGBoost algorithms, and also shows stochasticity due to the Gaussian noise used in the algorithm.
We therefore recommend to use WEvidential for better performance and efficient computation.
\Cref{fig:supplement_output} depicts the output of each algorithm with 100 weak learners trained.

\section{Additional Detail of Application} \label{apx:appendix_f}

This section describes additional details of the applications in \Cref{sec:section4}.
All the experiments were performed with x86-64 CPUs, where some of them were parallelised up to 10 CPUs and the rest uses 1 CPU.
The scripts to reproduce all the experiments are provided in the source code.
\Cref{apx:appendix_f1,apx:appendix_f2,apx:appendix_f3} describe additional details of the applications in, respectively, \Cref{sec:section41,sec:section42,sec:section43}.
\Cref{apx:appendix_f4} describes a choice of initial constants $\{ \vartheta_0^n \}_{n=1}^{N}$ for the WGBoost algorithm used in \Cref{sec:section4}.

\subsection{Detail of \Cref{sec:section41}} \label{apx:appendix_f1}

The normal response distribution $\mathcal{N}(y \mid m, \sigma)$ in \Cref{ex:normal} was used in \Cref{sec:section41}.
The normal output distribution has the scale parameter $\sigma$ that lies only in the positive domain of the Euclidean space $\R$.
We reparametrised it as one in the Euclidean space $\R$ by the log transform $\sigma' := \log(\sigma)$, which is the standard practice in Bayesian computation \cite{Gelman2013}.
The inverse of the log transform is the exponential transform $\sigma = \exp(\sigma')$.
It follows from change of variable that the individual-level posterior on $(m, \sigma')$ conditional on each individual observed response $y_i$, with the prior in \Cref{ex:normal}, is given by
\begin{align}
	\mu_i(m, \sigma') = p(m, \sigma' \mid y_i) \propto \exp\left( - \frac{1}{2} \frac{ ( y_i - m )^2 }{ \exp(\sigma')^2 } \right) \times \exp\left( - \frac{1}{2} \frac{ m^2 }{ 10^2 } \right) \times \frac{1}{\exp(\sigma')^{1.01}} \exp\left( - \frac{0.01}{\exp(\sigma')} \right)
\end{align}
up to the normalising constant, where we used the Jacobian determinant $| d \sigma / d \sigma' | = \exp(\sigma')$.

\subsection{Detail of \Cref{sec:section42}} \label{apx:appendix_f2}

The same reparametrisation of the normal output distribution as \Cref{apx:appendix_f2} was used in \Cref{sec:section42}.
For test data $\{ x_i, y_i \}_{i=1}^{D}$, the NLL and RMSE of each algorithm were computed by
\begin{align}
	\text{NLL} = - \frac{1}{D} \sum_{i=1}^{D} \log p(y_i \mid x_i) \quad \text{and} \quad \text{RMSE} = \sqrt{ \frac{1}{D} \sum_{i=1}^{D} (y_i - \hat{y}_i )^2 }
\end{align}
for the obtained provided predictive distribution $p(y_i \mid x_i)$ and the point prediction $\hat{y}_i$.
For WEvidential, the observed responses in training data were standardised. 
Accordingly, the test responses were standardised as $y_i' = ( y_i - y_{\text{mean}}^{\text{train}} ) / y_{\text{std}}^{\text{train}}$ using the mean $y_{\text{mean}}^{\text{train}}$ and standard deviation $y_{\text{std}}^{\text{train}}$ of the training responses.
The predictive distribution $p(y_i' \mid x_i)$ and point prediction $\hat{y}_i'$ of WEvidential were provided for the standardised responses $y_i'$.
The NLL and RMSE for the original responses $y_i$ can be computed as follows:
\begin{align}
	\text{NLL} & = - \frac{1}{D} \sum_{i=1}^{D} \log p(y_i \mid x_i) = - \frac{1}{D} \sum_{i=1}^{D} \log p(y_i' \mid x_i) + \log y_{\text{std}}^{\text{train}} , \\
	\text{RMSE} & = \sqrt{ \frac{1}{D} \sum_{i=1}^{D} (y_i - ( y_{\text{mean}}^{\text{train}} + y_{\text{std}}^{\text{train}} \times \hat{y}_i' ) )^2 } = y_{\text{std}}^{\text{train}} \sqrt{ \frac{1}{D} \sum_{i=1}^{D} ( y_i' - \hat{y}_i' )^2 } 
\end{align}
where the equality of the NLL follows from change of variable $p(y_i \mid x_i) = p(y_i' \mid x_i) / y_{\text{std}}^{\text{train}}$ and the equality of the RMSE follows from rearranging the terms.

We provide a brief description of each algorithm used for the comparison.
For each algorithm, the normal response distribution $p(y \mid m, \sigma)$ was specified for the response variable and the algorithm produces a point or distributional estimate of the response parameter $(m, \sigma)$ at each input $x$.

\begin{itemize}
	\item MCDropout \cite{Gal2016} trains a single neural network $F$ while dropping out each network weight with some Bernoulli probability. It can be interpreted as a variational approximation of a Bayesian neural network. It generates multiple subnetworks $\{ F^n \}_{n=1}^{N}$ by subsampling the network weights by the dropout. The predictive distribution $p(y \mid x)$ is given by the model averaging $(1 / N) \sum_{i=1}^{N} p(y \mid (m, \sigma) = F^n(x))$ for each input $x$.
	\item DEnsemble \cite{Lakshminarayanan2017} simply trains independent copies $\{ F^n \}_{n=1}^{N}$ of a neural network $F$ in parallel. It is one of the mainstream approaches to uncertainty quantification based on deep learning. The predictive distribution is given by the model averaging as in MCDropout.
	\item CDropout \cite{Gal2017} consider a continuous relaxation of the Bernoulli random variable used in MCDropout to optimise the Bernoulli probability of the dropout. It generates multiple subnetworks $\{ F^n \}_{n=1}^{N}$ by subsampling the network weights by the dropout with the optimised probability. The predictive distribution is the same as MCDropout.
	\item NGBoosting \cite{Duan2020} is a family of gradient booting that use the natural gradient \cite{Amari2016} of the response distribution as a target variable of each weak learner. In contrast to other methods, NGBoost outputs a single value $F(x)$ to be plugged into the response-distribution parameter. The predictive distribution $p(y \mid x)$ is given by $p(y \mid (m, \sigma) = F(x))$ for each input $x$.
	\item DEvidential \cite{Amini2020} extends deep evidential learning \cite{Sensoy2018}, originally proposed in classification settings, to regression settings. It considers the case where the individual-level posterior of the response distribution falls into a conjugate parametric form, and predicts the hyperparameter of the individual-level posterior by a neural network. The predictive distribution is also given in a conjugate closed-form.
\end{itemize}

\subsection{Detail of \Cref{sec:section43}} \label{apx:appendix_f3}

Similarly to the normal response distribution used in \Cref{sec:section41,sec:section42}, we reparametrised the parameter of the categorical response distribution used in \Cref{sec:section43}.
The categorical response distribution $\mathcal{C}(y \mid q)$ in \Cref{ex:categorical} has a class probability parameter $q = [ q_1, \dots, q_k ]$ in the simplex $\Delta_k$.
We reparametrised the parameter $q$ by the log-ratio transform $q' := [ \log( q_1 / q_{k} ), \dots, \log( q_{k-1} / q_{k} ) ] \in \R^{k-1}$ that maps from the simplex $\Delta_k$ to the Euclidean space $\R^{k-1}$ \cite{Aitchison1980}.
The inverse is the logistic transform 
\begin{align}
	q = \left[ \frac{ \exp( q'_1 ) }{ z_k } , \dots, \frac{ \exp( q'_{k-1} ) }{ z_k }, \frac{ 1 }{ z_k } \right] \in \Delta_k \quad \text{where} \quad z_k = 1 + \sum_{j=1}^{k-1} \exp( q'_j ) .
\end{align}
The logistic normal distribution on the original parameter $q$ corresponds to a normal distribution on the transformed parameter $q'$ by change of variable \cite{Aitchison1980}.
By change of variable, the individual-level posterior on $q'$ conditional on each individual observed response $y_i$, with the prior in \Cref{ex:categorical}, is
\begin{align}
	\mu_i(q') = p(q' \mid y_i) \propto \left( \frac{ 1 }{ z_k } \right)^{ [y_i = k] } \times \prod_{j=1}^{k-1} \left( \frac{ \exp( q'_j ) }{ z_k } \right)^{ [y_i = j] } \times \exp\left( - \frac{1}{2} \frac{ \| q' \|^2 }{ 10^2 } \right)
\end{align}
up to the normalising constant, where $[y_i = j]$ is $1$ if $y_i$ is the $j$-th class label and $0$ otherwise.

We provide a brief description of each algorithm used in comparison with WGBoost.
MCDropout and DEnsemble are described in \Cref{apx:appendix_f2}.

\begin{itemize}
	\item DDistillation \cite{Malinin2020} predicts the parameter of a Dirichlet distribution over the simplex $\Delta_k$ by a neural network using the output of DEnsemble. The output of multiple networks in DEnsemble is distilled into the a Dirichlet distribution controlled by one single network.
	\item PNetwork \cite{Charpentier2020} considers the case where the individual-level posterior of the categorical response distribution falls into a Dirichlet distribution similarly to deep evidential learning \cite{Sensoy2018}. It predicts the hyperparameter of the individual-level posterior given in the form of the Dirichlet distribution.
\end{itemize}

\subsection{Choice of Initial State of WGBoost} \label{apx:appendix_f4}

In standard gradient boosting, the initial state at step $m = 0$ is specified by a constant that most fits given data.
Similarly, we specified the initial state $\{ \vartheta_0^n \}_{n=1}^{N}$ of WEvidential by a set of constants that most fits the output distributions in average.
We find such a set of constants by performing an approximate Wasserstein gradient flow averaged over all the output distributions.
Specifically, given the term $\G_i(\mu)$ in Algorithm \ref{alg:wgb}, we define $\bar{\G}(\mu) := (1 / D) \sum_{i=1}^{D} \G_i(\mu)$ and perform the update scheme of a set of $N$ particles $\{ \bar{\vartheta}_m^n \}_{n=1}^{N}$:
\begin{align}
	\begin{bmatrix}
		\bar{\vartheta}_{m+1}^1 \\
		\vdots \\
		\bar{\vartheta}_{m+1}^N
	\end{bmatrix} 
	= 
	\begin{bmatrix}
		\bar{\vartheta}_m^1 \\
		\vdots \\
		\bar{\vartheta}_m^N
	\end{bmatrix}
	+ \nu_0
	\begin{bmatrix}
		- [ \bar{\G}(\hat{\pi}_m) ]( \bar{\vartheta}_m^1 ) \\
		\vdots \\
		- [ \bar{\G}(\hat{\pi}_m) ]( \bar{\vartheta}_m^N )
	\end{bmatrix} \label{eq:initial_particle_update}
\end{align}
with the learning rate $\nu_0 = 0.01$ up to the maximum step number $m = 5000$.
The initial particle locations for this update scheme were sampled from a standard normal distribution.
We specified the initial state $\{ \vartheta_0^n \}_{n=1}^{N}$ by the set of particles $\{ \bar{\vartheta}_{m}^n \}_{n=1}^{N}$ obtained though this scheme at $m = 5000$.